\documentclass[sigconf,screen]{acmart}

\newcommand{\singlecolumnwidth}{\linewidth}

\AtBeginDocument{%
  }

\setcopyright{
  acmlicensed
}

\copyrightyear{2024}
\acmYear{2024}
\acmDOI{https://doi.org/10.1145/3613904.3642944}

\acmConference[CHI '24]{the 2024 CHI Conference on Human Factors in Computing Systems}{May 11-16,
  2024}{Honolulu, Hawai`i}
\acmISBN{}
\usepackage{algorithm}
\usepackage{wrapfig}
\usepackage{multirow}

\usepackage[indLines=true,noEnd=true]{algpseudocodex}
\DeclareDocumentCommand{\todo}{o}{
  \textcolor{red}{
    TODO\IfValueT{#1}{: #1}
  }
}

\newcommand{\bs}[1]{\textbf{\textsf{#1}}}

\newcommand{\yhydone}[1]{#1}

\newcommand{\hide}[1]{}

\newcommand{\revision}[1]{#1}
\newcommand{\revisiontext}[1]{#1}
\newcommand{\revisionbox}[1]{#1}
\newcommand{\sidecomment}[1]{}

\begin{document}

\title[SalienTime: User-driven Salient Time Steps Selection for Geospatial Visualization]{SalienTime: User-driven Selection of Salient Time Steps for Large-Scale Geospatial Data Visualization}


\author{Juntong Chen}
\email{jtchen@stu.ecnu.edu.cn}
\orcid{0000-0001-9343-4032}
\affiliation{%
  \institution{School of Computer Science and Technology\\East China Normal University}
  \city{Shanghai}
  \country{China}
}

\author{Haiwen Huang}
\email{hwhuang@stu.ecnu.edu.cn}
\orcid{0009-0005-9443-3855}
\affiliation{%
  \institution{School of Computer Science and Technology\\East China Normal University}
  \city{Shanghai}
  \country{China}
}

\author{Huayuan Ye}
\email{huayuan221@gmail.com}
\orcid{0009-0008-8208-2017}
\affiliation{%
  \institution{School of Computer Science and Technology\\East China Normal University}
  \city{Shanghai}
  \country{China}
}

\author{Zhong Peng}
\email{zpeng@sklec.ecnu.edu.cn}
\orcid{0000-0003-2033-9258}
\affiliation{%
  \institution{State Key Laboratory of Estuarine and Coastal Research\\East China Normal University}
  \city{Shanghai}
  \country{China}
}

\author{Chenhui Li}
\email{chli@cs.ecnu.edu.cn}
\orcid{0000-0001-9835-2650}
\affiliation{%
  \institution{School of Computer Science and Technology\\East China Normal University}
  \city{Shanghai}
  \country{China}
}

\author{Changbo Wang}
\email{cbwang@cs.ecnu.edu.cn}
\orcid{0000-0001-8940-6418}
\affiliation{%
  \institution{School of Computer Science and Technology\\East China Normal University}
  \city{Shanghai}
  \country{China}
  % \postcode{200063}
}

% \affiliation{%
%   \instituti on{Institute for Clarity in Documentation}

% \affiliation{%



% \renewcommand{\shortauthors}{Chen et al.}

\begin{abstract}

The voluminous nature of geospatial temporal data from physical monitors and simulation models poses challenges to efficient data access, often resulting in cumbersome temporal selection experiences in web-based data portals. Thus, selecting a subset of time steps for prioritized visualization and pre-loading is highly desirable. Addressing this issue, this paper establishes a multifaceted definition of salient time steps via extensive need-finding studies with domain experts to understand their workflows. Building on this, we propose a novel approach that leverages autoencoders and dynamic programming to facilitate user-driven temporal selections. Structural features, statistical variations, and distance penalties are incorporated to make more flexible selections. User-specified priorities, spatial regions, and aggregations are used to combine different perspectives. \revision{We design and implement a web-based interface to enable efficient and context-aware selection of time steps and evaluate its efficacy and usability through case studies, quantitative evaluations, and expert interviews.
%on diverse real-world geospatial datasets.
}

\end{abstract}

% \ccsdesc[300]{Human-centered computing~Interaction design process and methods}
% \ccsdesc[300]{Theory of computation~Dynamic programming}




\begin{teaserfigure}
  \centering
  \includegraphics[width=0.95\textwidth]{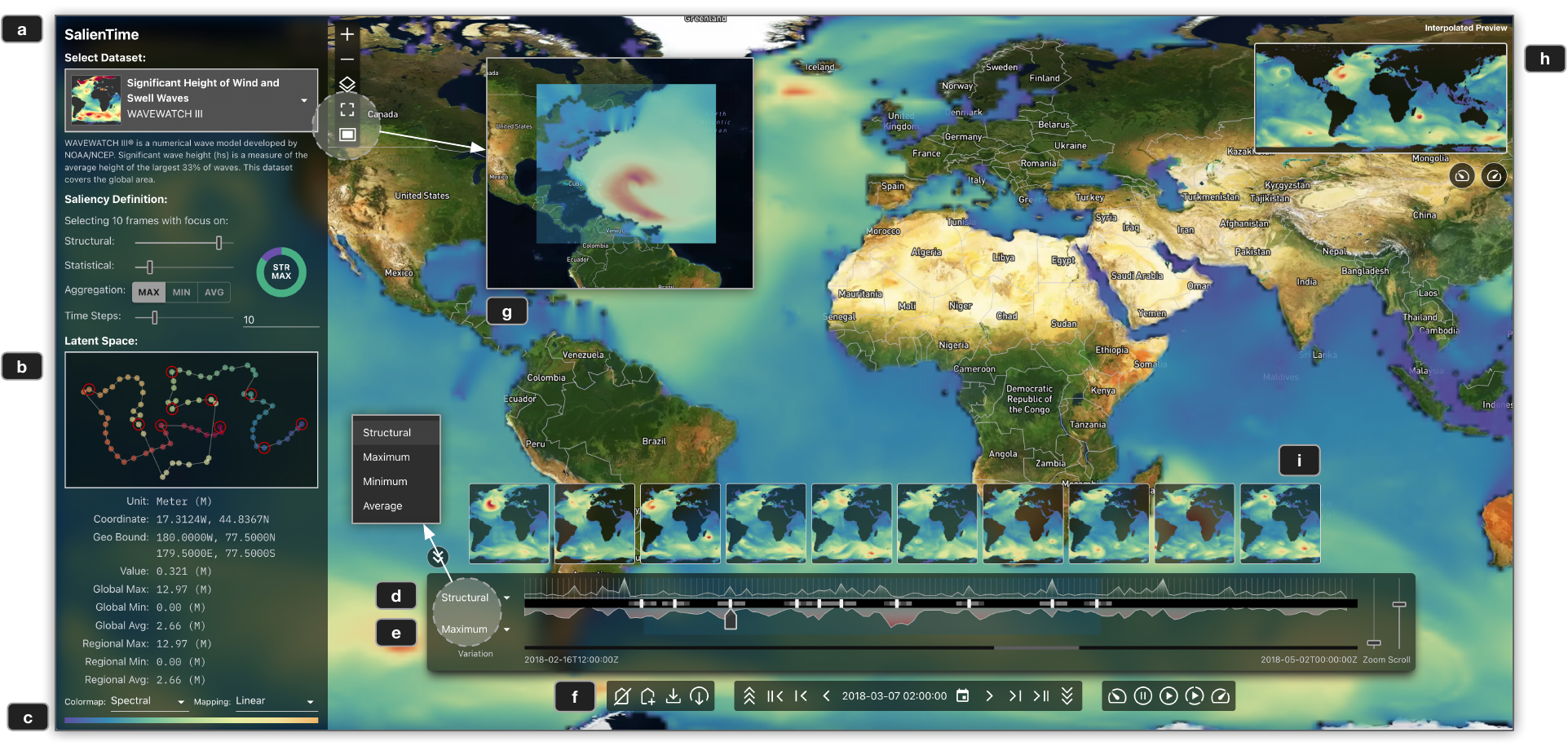}
  \caption{SalienTime User Interface. The dataset displayed here is the significant wave height of a 2-year global wave hindcast data. The contextual visualization we introduced (\bs{d}, \bs{e}) and the latent space exploration (\bs{b}) facilitate efficient exploration and selection of salient time steps from large-scale geospatial data. More details are elaborated in Section \ref{sec:system}.}
  \label{fig:teaser}
  \Description{
  A screenshot for the SalienTime system, where a colorized heat map for significant wave height is layered on a satellite map. Letters mark the UI components. On the left side (from top to bottom):
  a. Dataset selection and parameter configuration;
  b. Latent space visualization;
  c. Colormap and mapping function configuration;
  In the middle:
  d. Temporal trend visualization;
  e. Relative trend visualization;
  f. Function panel;
  g. Spatial boundary selection;
  i. Salient frames.
  On the right:
  h. Playback view;
  }
\end{teaserfigure}

\maketitle

\section{Introduction}

% \begin{figure*}[htbp]
%     \includegraphics[width=\textwidth]{figs/teaser.eps}
%     \caption{SalienTime User Interface. The dataset displayed here is the significant wave height of a 2-year global wave hindcast data. The contextual visualization we introduced (\bs{d}, \bs{e}) and the latent space exploration (\bs{b}) facilitate efficient exploration and selection of salient time steps from large-scale geospatial data. More details are elaborated in Section \ref{sec:system}.}
%     \label{fig:teaser}
%     \Description{
%     A screenshot for the SalienTime system, where a colorized heat map for significant wave height is layered on a satellite map. Letters mark the UI components. On the left side (from top to bottom):
%     a. Dataset selection and parameter configuration;
%     b. Latent space visualization;
%     c. Colormap and mapping function configuration;
%     In the middle:
%     d. Temporal trend visualization;
%     e. Relative trend visualization;
%     f. Function panel;
%     g. Spatial boundary selection;
%     i. Salient frames.
%     On the right:
%     h. Playback view;
%     }
% \end{figure*}

% Geopsatial data’s definition. I want to focus on earth observation data collected on satellites, remote sensing images, and model simulation data and use figures to show that they are really big data.
With the rapid advancements in physical monitoring technologies and numerical simulation models, geospatial data has grown exponentially in both ubiquity and volume ~\cite{yangGeospatialCyberinfrastructurePresent2010}. Today, there are over 1000 Earth Observation satellites orbiting the globe, producing terabytes to petabytes of observation data daily. A single full-resolution remote-sensing image can take a few hundred megabytes of storage space  ~\cite{yaoEnablingBigEarth2019}. Numerical hindcast models such as WAVEWATCH III ~\cite{tolmanUserManualSystem2014} output gigabytes of global oceanography data hourly. This wealth of data serves as a powerful tool to analyze natural phenomena and geographical characteristics such as water quality assessment, natural disaster monitoring, climate change detection, land use mapping, etc. ~\cite{dengGeospatialBigData2019} As the spatial resolution increases and more satellites launch missions on the horizon, the data volume will only get larger in the future ~\cite{leeGeospatialBigData2015}, bringing new challenges for storing, accessing, processing, and visualization.

The true potential of these geospatial big data is only realized when accessible to users, necessitating an effective approach for data access. As of 2022, the Sentinel project ~\cite{druschSentinel2ESAOptical2012} by the European Space Agency (ESA) has published over 60 million datasets, gaining more than 500 PB of downloads since the launch of the operation ~\cite{copernicusSentinelData2022}.
% Earth Observing System Data and Information System (EOSDIS) Project ~\cite{EOSDISAnnualMetrics2023} by NASA has distributed over 100 PB of products in 2022.
Currently, the typical data infrastructure for serving these geospatial data is to store them in a cloud-based storage system~\cite{yangUtilizingCloudComputing2017}, organized by spatial resolution and temporal granularity, allowing users to access via authenticated HTTP/FTP servers, or APIs. In the traditional workflow, researchers often faced lengthy periods to download the full-sized data to the local workstation before performing analytical tasks. A recent shift towards more user-friendly approaches has been observed. Many organizations now offer a data access hub: a map-centered web platform that facilitates interactive data exploration and visualization directly on maps. Notable examples of such platforms include NASA Worldview ~\cite{NASAWorldview2023}, NOAA View Global Data Explorer ~\cite{NOAAViewGlobal2023}, USGS EarthExplorer ~\cite{USGSEarthExplorer2023}, Copernicus Open Access Hub ~\cite{CopernicusOpenAccess2023}, etc. One common feature of these platforms is to create time-lapse playbacks ~\cite{SentinelHubEO2023, NASAWorldview2023}, allowing users to observe the temporal variation of the data. In such workflows, users begin by choosing a dataset, then visualize a segment of the data to determine its appropriateness. After confirmation, they specify regional and temporal constraints to download the full dataset. With these data exploration platforms, users are empowered to navigate efficiently and assess the dataset's usability without extensive downloads.

However, due to limited network bandwidth and the voluminous nature of geospatial data, visualizing a single frame often demands several seconds. When users switch to a different time step, they confront another waiting period. Though platforms like NASA Worldview have attempted to address this issue with buffer queues, which pre-downloads subsequent data when users inspect data at a certain time step, temporal data exploration remains cumbersome. Moreover, users often navigate between time steps randomly without adequate contextual information, leaving them in the dark about potential patterns and correlations. Ziegler et al. ~\cite{zieglerNeedFindingStudyUsers2023} identified it as one of the major challenges in geospatial systems: users struggle to find desired temporal constraints for data. Therefore, if a representative subset of time steps from the original dataset can be selected for prioritized download and visualization, users can swiftly refine their time selections and identify desirable download targets efficiently.

This problem defines the scope of our research: selecting an informative subset of time steps from large-scale geospatial temporal data to assist users with temporal selection. Existing research primarily aims to minimize the reconstruction error of the \textit{piece-wise linear interpolated} ~\cite{wuStreamingApproachSitu2022} data as the goal for selection. That is, the data between two selected time steps are linearly interpolated to generate full-duration data as the reconstructed data for comparison. To achieve this, dynamic programming-based methods ~\cite{tongSalientTimeSteps2012, zhouKeyTimeSteps2018} and deep learning-based methods ~\cite{porterDeepLearningApproach2019, liuKeyTimeSteps2019} are two main approaches. These methods underscore the \textit{summerizability} of the selected time steps, employing metrics like RMSE and SSIM to quantify the quality of the selection. However, our need-finding study shows that under different datasets and analytics tasks, users have different requirements for time step selection. Users are not solely interested in the summerizability of the selected time steps. They also value the amount of information brought by the subset of data in different aspects. Moreover, they anticipate supplemental contexts to guide their selection. Current research that primarily focuses on reconstruction error does not apply to these specific needs, and a user-driven approach that allows users to define priorities for time selection is highly desirable.

In this paper, we first establish a multifaceted definition for salient time steps in the context of geospatial big data via an extensive need-finding study with domain exports. We then introduce a novel approach that leverages autoencoders and Dynamic Programming (DP) algorithm to facilitate user-driven selections of salient time steps. Our network is designed to capture the intrinsic structure of the data, generating meaningful latent codes for measuring structural similarity. Statistical variation and distance penalty are incorporated in the DP process to make more flexible selections. Different perspectives of selection are combined with users-specified priorities, spatial regions and aggregation methods. Additionally, we design and implement an interactive system that enables users to efficiently identify salient time steps from large-scale geospatial data. The efficacy of our approach is demonstrated via case studies on algal bloom and hurricane data and quantitative evaluations on 5 diverse geospatial datasets.

% According to the International Data Corporation (IDC), the volume of geospatial data is growing at a rate of 40\% per year, and the total volume of geospatial data is expected to reach 175 Zettabytes by 2025 @CITE. With the increasing spatial resolution, the data will only gets larger in the future.

% In the rest of the paper, we use \textit{frame} to refer to a single time step of the data, and \textit{salient frames} to the subset of time steps we selected from the full data. We use \textit{from the data}...

In summary, the contributions of this paper are:

\begin{itemize}
    \item We establish a multifaceted definition for salient time steps in the context of geospatial big data via an extensive need-finding study with domain exports.
    \revision{\sidecomment{R3.1\\MR1}
    \item We propose a novel approach that employs autoencoders and dynamic programming to select salient time steps from geospatial data with user-specified priorities. We incorporate our approach in an interactive system, facilitating efficient temporal navigation and data exploration.
    \item We conduct two case studies, quantitative evaluations on diverse real-world datasets and expert interviews to evaluate the efficacy of our approach and the usability of our system.
    }
\end{itemize}

% \section{Background \& Related Work}

\section{Related Work}

% \todo[Background - Geospatial Big Data]

% \todo[User-driven approach / User defined conditions]

% \todo[Time Selection of spatiotemporal Data]

% \todo[Encoder-decoder Networks]

% Background - Big Data in GIS

% User-driven approach / used defined conditions

% Selecting timesteps. (Three paragraphs)

\subsection{Time Selection from Spatiotemporal Data}

% \myworries{unpolished}

Selecting a subset of representative time steps automatically from temporal data has been an active field of research in scientific visualization. Judicious selection of salient time steps can lower the network bandwidth requirement and facilitate rendering time allocation ~\cite{chaoliwangImportanceDrivenTimeVaryingData2008}. One approach is clustering, where Akiba et al. ~\cite{akibaSimultaneousClassificationTimeVarying2006} group similar time steps into clusters and select one time step from each group. However, this fails to capture the chronological information and may lead to repeated selection clustered in a short period. Another prevalent approach is to employ specific metrics to quantify the variation between time steps and select frames that are most dissimilar from their predecessors. Common metrics include entropy-based measures, where Wang et. al ~\cite{chaoliwangImportanceDrivenTimeVaryingData2008} employ mutual information, Zhou et al. ~\cite{zhouKeyTimeSteps2018} and Dutta et al. ~\cite{duttaSituAdaptiveSpatioTemporal2021} use \textit{variation of information}, which is based on conditional entropy. Tong et al. ~\cite{tongSalientTimeSteps2012} use Dynamic Time Wrapping (DTW), which measures the distance between two sequences to map the entire sequence onto a user-specified number of key time steps. In terms of algorithms, greedy-based methods ~\cite{woodringMultiscaleTimeActivity2009,chaoliwangImportanceDrivenTimeVaryingData2008,wuStreamingApproachSitu2022} and dynamic programming-based methods ~\cite{tongSalientTimeSteps2012, zhouKeyTimeSteps2018} are commonly employed. The greedy approach considers local variation and looks for maximizing the variation of the current time step, thus having no optimality guarantees ~\cite{wuStreamingApproachSitu2022}. Dynamic programming-based methods consider all combinations and are guaranteed to find the global optimal results.
% However, the computation needed is higher for the worst time complexity of $O(n^3)$.
Zhou et al. ~\cite{zhouKeyTimeSteps2018} propose an approximation approach, in which a multi-pass algorithm with sliding windows is employed to reduce the computation needed for the cost function.
% Other techniques include Frey et al.'s flow-based method ~\cite{freyFlowBasedTemporalSelection2017} and Pulido et al.'s tensor decomposition approach ~\cite{pulidoSelectionOptimalSalient2021}.
Other techniques include a flow-based method proposed by Frey et al. ~\cite{freyFlowBasedTemporalSelection2017}, starting with random samples and employing minimum-cost flow to progressively adjust time steps that maximize the coverage with the original data. Pulido et al. ~\cite{pulidoSelectionOptimalSalient2021} applies non-negative Tucker tensor decomposition to extract latent time features and map these features to their maximal values.

Recently, another branch of deep learning-based method has been proposed. Porter et al. ~\cite{porterDeepLearningApproach2019} uses an Autoencoder network to encode each time step to latent space and use a combination of arc-length-based and angle-based selection to choose frames in the dimension-reduced latent space. In the same year, Liu et al. ~\cite{liuKeyTimeSteps2019} proposed to employ Deep Metric Learning (DML) that utilizes a Siamese deep neural network to select key time steps in a supervised way.

\revision{\sidecomment{R1.1\\MR6}
To evaluate the quality of selection results, one common approach is to reconstruct data from the selected time steps via linear interpolation and quantify total reconstruction errors. This is typically carried out using image quality metrics such as Root Mean Square Error (RMSE) \cite{zhouKeyTimeSteps2018, pulidoSelectionOptimalSalient2021,porterDeepLearningApproach2019}, Peak Signal-to-Noise Ratio (PSNR) \cite{pulidoSelectionOptimalSalient2021,porterDeepLearningApproach2019}, Structural Similarity Index Measures (SSIM) ~\cite{pulidoSelectionOptimalSalient2021}, and Total Absolute Error (TAE) ~\cite{pulidoSelectionOptimalSalient2021}. Another approach is to directly evaluate the cost metric used for frame selection, such as the Dynamic Time Wrapping distance by Tong et al.~\cite{tongSalientTimeSteps2012}, and the Flow-based distance by Frey et al.~\cite{freyFlowBasedTemporalSelection2017}. In our study, we use linear interpolation as an intuitive method to reconstruct the full data, and aligning previous studies, employ RMSE and SSIM to measure reconstruction quality.
}

Leveraging the feature representation ability of neural networks, deep learning-based methods can effectively capture the structural variations in the data. Meanwhile, DP-based methods offer high explainability, flexible cost definitions and global optimality. Our approach combines the two, selecting informative salient time steps that address user needs.

% Woodring et al. ~\cite{woodringMultiscaleTimeActivity2009} utilize wavelet transform to provide multi-scale temporal exploration and time step selection.

\subsection{User-driven Approaches for Geospatial Visualization Systems}
% This section sucks
% \myworries{unpolished}

\revision{
\sidecomment{R3.5\\MR5}
Visualization systems for geospatial data are an active field of research due to their significance in helping researchers understand and analyze the pattern, trend, uncertainty and correlation of data \cite{chenSurveyMultiSpaceTechniques2019}. Numerous systems are proposed for various applications, such as climatology\cite{dengAirVisVisualAnalytics2019}, aerodynamics\cite{hanFlowNetDeepLearning2020}, social media analysis\cite{knittelRealTimeVisualAnalysis2022}, and urban computing \cite{shenStreetVizorVisualExploration2018, ChenSenseMapUrbanPerformance2023}, as comprehensively reviewed in surveys by Chen et al. \cite{chenSurveyMultiSpaceTechniques2019} and Deng et al. \cite{dengSurveyUrbanVisual2023}
}However, bias in the understanding of users' needs for spatial and temporal selection in geospatial systems is widespread and challenging to address~\cite{zieglerNeedFindingStudyUsers2023, traynorWhyAreGeographic1995} The concept \textit{user-driven approaches} in geospatial systems encompasses quite a broad scope. It refers to the idea that we allow users to define their specified constraints or priorities to perform selections, rather than making presumptions about their preferences. For instance, MixMap~\cite{mckenzieMixMapUserdrivenApproach2023} focuses on place-based similarity, allowing users to assign weights to different regional characteristics like age, education, and race, to identify similar regions. Zhou et al. ~\cite{zhouUserDrivenSamplingModel2023} introduce a user-driven sampling model where users select representative points. A CNN network then learns these preferences to guide sampling across other spatial regions. \revision{\sidecomment{R1.1c}Liu et al. propose TPFlow ~\cite{liuTPFlowProgressivePartition2019} that allow users to select informative spatiotemporal slices from large-scale spatiotemporal data. Based on tensor decomposition, users can progressively drill down different data dimensions to gain insight.} The term also implies that access to spatial data should align with users' interactions and activity history ~\cite{stockhauseUserDrivenData2013}. For instance, Guo et al. ~\cite{guoEfficientSelectionGeospatial2018a} samples a subset of spatial objects based on different zoom levels and panning locations, facilitating efficient selection from large-scale geospatial data. Pan et al. ~\cite{panGlobalUserDrivenModel2017} leverages collective user behaviors to predict which tiles should be prefetched or replaced in the cache to improve geospatial system performance. In terms of selecting salient time steps, \textit{user-driven} also refers to providing contextual information with visualizations to assist interactive temporal selections, following the focus + context visualization scheme ~\cite{shneidermanEyesHaveIt1996}. For example, TransGraph ~\cite{yiguTransGraphHierarchicalExploration2011} provides hierarchical visualizations of the time steps, allowing users to select time steps at different granularity levels. Dissimilarity maps ~\cite{tongSalientTimeSteps2012} are 2D heatmaps that illustrate the differences between the current selection and the original data. Time Activity Curves (TACs) ~\cite{teng-yokleeVisualizingTimevaryingFeatures2009,teng-yokleeVisualizationExplorationTemporal2009, woodringMultiscaleTimeActivity2009} provide visualizations about the magnitude of variation, offering insights into temporal data patterns.

\revision{\sidecomment{R1.1b}
We adopt the aforementioned user-driven ideologies by offering users adjustable priorities for selection, ensuring our approach applies to different tasks. We also incorporated contextual visualizations, including temporal and relative trends, to assist users in refining their current selection. Additionally, our system takes into account the user's previous manual labeling when computing selections.
}

% In the context of selecting time steps,  TransGraph ~\cite{yiguTransGraphHierarchicalExploration2011}, Time Activity Curve ~\cite{teng-yokleeVisualizationExplorationTemporal2009}

% GAN disentanglement ~\cite{evirgenGANravelUserDrivenDirection2023}
% Layout ~\cite{niyazovUserDrivenConstraintsLayout2023}

\subsection{Spatial Feature Representation}

Feature representation plays an important role in various domains, such as dimensionality reduction ~\cite{joliffe1992principal, maatenVisualizingDataUsing2008, 2018arXivUMAP}, anomaly detection ~\cite{erfani2016high, ionescu2019object}, data steganography~\cite{zhangVisCodeEmbeddingInformation2020, yeInvVisLargeScaleData2023}, etc. Early approaches are commonly based on traditional machine learning methods. Principal Component Analysis (PCA) ~\cite{joliffe1992principal} projects the original data onto a low-dimensional space, retaining the main features of data variation. Linear Discriminant Analysis (LDA) ~\cite{izenman2013linear} uses a supervised learning method to find a linear combination of features that characterizes different classes of objects. Papadimitriou et al.~\cite{Papadimitriou1998LatentSI} proposed to leverage Singular Value Decomposition (SVD) to implement feature extraction, which is later widely used in recommender system ~\cite{Koren2009MatrixFT, Paterek2007ImprovingRS}. There are also some feature representation methods based on domain transform, e.g., Wavelet Transform ~\cite{Soro2019AWS} and Fourier Transform ~\cite{Singh2016FourierBasedFE}.

Recently, a variety of deep learning-based feature representation schemes have been proposed and achieve impressive performance. These methods are generally based on autoencoders. Autoencoder is a kind of neural network that is designed to efficiently learn a low-dimensional representation (which is usually called latent code) for a set of unlabelled data. The learned latent code can be used for numerous kinds of downstream tasks, such as image reconstruction and representation ~\cite{Hinton2006ReducingTD, Vincent2008ExtractingAC,jiangLatentMapEffectiveAutoencoding2021}, sentiment prediction ~\cite{Tang2014LearningSW}, etc. There are some studies ~\cite{Goodfellow2014GenerativeAN, Radford2015UnsupervisedRL,huangDeSmoothGANRecoveringDetails2020} that combine autoencoder and adversarial training to implement image generation, which is generating new images using the learned latent code representation. Kingma et al. proposed Variational AutoEncoder (VAE) ~\cite{Kingma2013AutoEncodingVB} for generation tasks by adding conditional constraints on latent code.

In this paper, we propose a new method utilizing autoencoders to efficiently learn feature representation of spatiotemporal data and quantify the structural variation of data frames based on the learned latent code.
\section{Need-finding Study}
\label{sec:needfinding}

\begin{figure*}[tbhp]
    \centering
    \includegraphics[width=\linewidth]{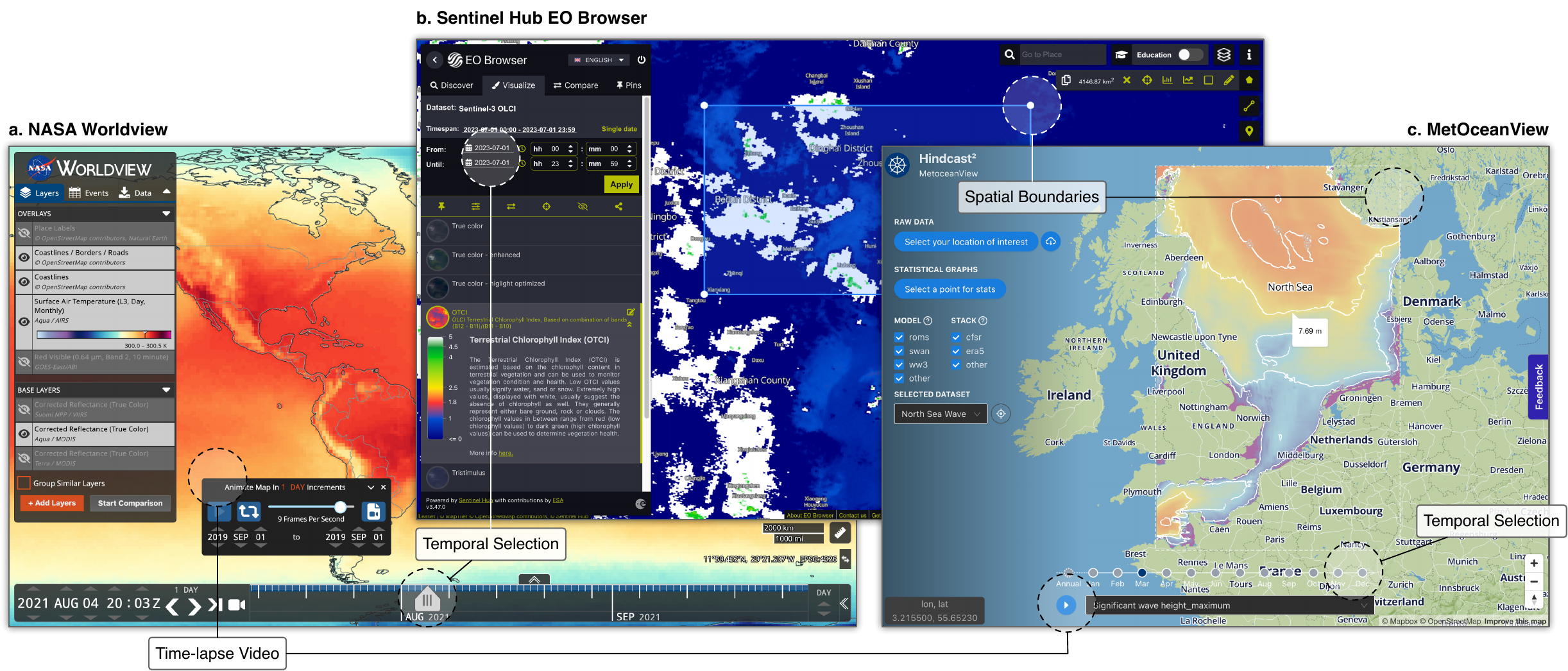}
    \caption{
        The platforms ~\cite{NASAWorldview2023, SentinelHubEO2023, MetOceanView2023} we used for the observational study.
    }
    \label{fig:platforms}
    \Description{Three screenshots for geospatial data access platforms. From left to right are NASA Worldview, Sentinel Hub EO Browser, and MetOceanView. The UI components for creating time-lapse videos, and selecting temporal and spatial constraints are highlighted with circles.}
\end{figure*}

To gain insights into what characteristics make time steps \textit{salient} to users and require prioritized visualization, we conducted a need-finding study with domain experts through observational studies and semi-structured interviews. In this section, we first describe the participants and structure of our interview, then elaborate on our findings.

\subsection{Methodology}

\subsubsection{Participants} We engaged with 5 domain experts from both academia and industry (age $\mu = 32.6$, $\sigma=6.8$) via email contact, direct messages and verbal communication. Expert 1 (E1) has held the position of Principal Oceanographer in a research institute for more than five years, and has transitioned into academia, being a professor specializing in Coastal Dynamics at University E. Expert 2 (E2) is a senior engineer in an ecological institute, currently in charge of a project about algal bloom monitoring and early warning. Expert 3 (E3) is a software engineer working in the data visualization department of a leading company. Expert 4 (E4) is a professor at University N, studying urban data visualization, pervasive computing, and human-computer interaction. Expert 5 (E5) is an associate professor at University E specializing in data visualization and spatiotemporal data mining. The participants' domains and backgrounds are summarized in \autoref{tab:expert}.

\begin{table}[hbtp]
    \caption{Domain and experience of the experts in our need-finding study.}
    \small
    \label{tab:expert}
      \begin{tabular}{llll}
          \toprule
          ID & Domain             & Background & Years of Exp.   \\
          \midrule
          E1 & Earth Science      & Academia & \textgreater{}10  \\
          E2 & Earth Science      & Industry & 5-10              \\
          E3 & Data Visualization & Industry & 5-10              \\
          E4 & Data Visualization & Academia & 5-10              \\
          E5 & Data Visualization & Academia & \textgreater{}10  \\
      \bottomrule
      \end{tabular}%
  \end{table}

\subsubsection{Platforms for Observation} To observe the participant's usage pattern for temporal selection within geospatial systems, we use three widely-used platforms: NASA Worldview ~\cite{NASAWorldview2023}, Sentinel Hub EO Browser ~\cite{SentinelHubEO2023} and MetOceanView ~\cite{MetOceanView2023}.
\revision{\sidecomment{R1.1a}These platforms are selected as they are frequently used by geoscience and visualization researchers to retrieve data, as suggested by E1, E2 and E5. They also exhibit similar functionality.} NASA Worldview enables real-time interaction with over 1,000 global, full-resolution satellite imagery layers, updated hourly to support time-critical applications. Sentinel Hub EO Browser provides a complete archive of Sentinel, Landsat, and other ESA satellite products, with customizable visualizations and highlight recommendations for earth events. MetoceanView specializes in oceanography hindcast and forecast data, allowing animated visualizations across hundreds of datasets and models. \autoref{fig:platforms} shows screenshots of these platforms.

These platforms exhibit three core commonalities: 1). They serve as the web-based access portal to an extensive amount of data; 2). They provide categorical, spatial and temporal constraints to facilitate dataset discovery; 3). They offer map-based visualizations and time-lapse video creation capabilities. Users will undergo multiple rounds of filter, exploring and visualization iterations to find the download target, allowing us to understand the decision-making process involved.

\subsubsection{Session Structure} Each session was conducted in a face-to-face setting, lasting about 60 minutes. The session starts with a 10-minute introduction that outlines the research context and objectives. This is followed by a 25-minute free exploration phase and a 25-minute semi-structured interview. During the introduction phase, we asked for the participants' consent about screen recording for further analysis. In the exploration phase, we observed participants' behavior and encouraged them to describe their ongoing intentions and task status.\revision{\sidecomment{R1.2b} They were first instructed to explore all three platforms' functionality in terms of dataset selection, spatial and temporal navigation, and time-lapse video creation and then perform at least one full data utilizing insights provided by previous actions.
%   While there were no explicit directives on which dataset to download, participants were required to perform at least one full data download
We do provide explicit directives on which dataset to download due to the participants' different backgrounds and domain knowledge.} Allowing participants to choose their tasks also helps us to get a more authentic understanding of real-world conditions and more diverse insights.

The interview phase is structured around the following outline:

\begin{enumerate}
    \item \textbf{Motivation \& Outcome}: What dataset and the time range were selected, and what motivated the selection of the dataset and time range?
    \item \textbf{Decision-making Factors}: What contextual information from the platforms influenced their decisions in terms of spatial and temporal constraints when exploring datasets?
    \item \textbf{Temporal Selection}: How do they navigate between time steps for visualization, and how is the selection of the time range for download made?
    \item \textbf{Interaction Remarks}: What were the participants' assessments of the interactions involved within the data exploration and visualization workflow?
\end{enumerate}

We also incorporate our observations and may revisit screen recordings based on user feedback. We concurrently take notes to capture key insights.

% \subsection{Observations}
%
% \subsection{Findings}

\subsection{Findings}

\begin{figure*}[hbtp]
    \centering
    \includegraphics[width=\linewidth]{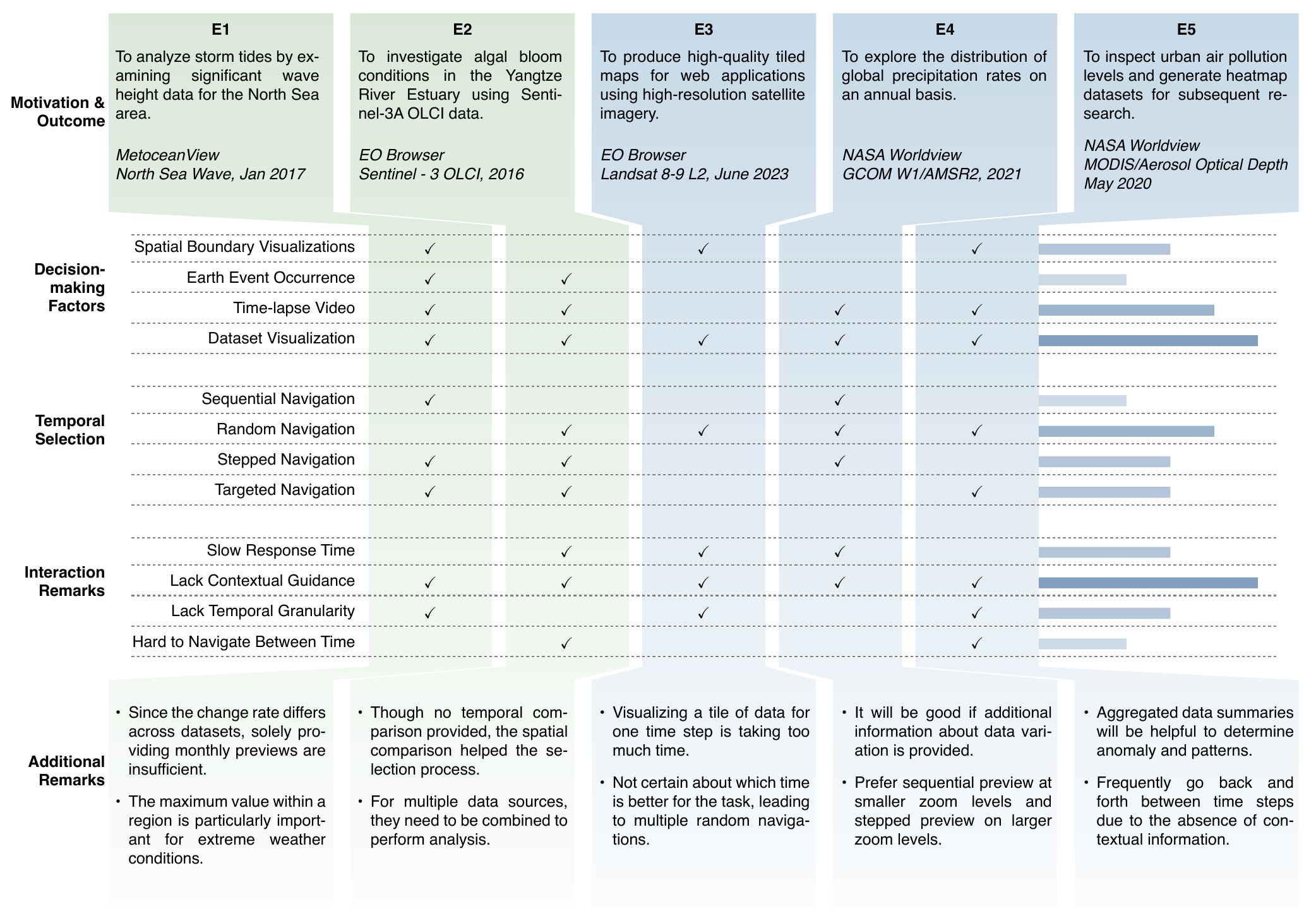}
    \caption{The compiled results of our need-finding study. The participants' domains are illustrated in backgrounds with green (Earth Science) or blue (Data Visualization) color.}
    \label{fig:findings}
    \Description{
A five-column, five-row data table with visualizations. The middle three rows contain multiple checks indicating whether the expert has mentioned the corresponding item.

E1: Motivation: To analyze storm tides by examining significant wave height data for the North Sea area; Outcome: MetoceanView - North Sea Wave, Jan 2017; Additional Remarks: "Since the rate of change differs across datasets, solely providing monthly previews are insufficient; The maximum value within a region is particularly important for extreme weather conditions."

E2: Motivation: To investigate algal bloom conditions in the Yangtze River Estuary using Sentinel-3A OLCI data; Outcome: EO Browser - Sentinel-3 OLCI, 2016; Additional Remarks: "Though no temporal comparison provided, the spatial comparison helped the selection process; Multiple data sources needs to be combined to perform analysis."

E3: Motivation: To produce high-quality tiled maps using high-resolution satellite imagery; Outcome: EO Browser - Landsat 8-9 L2, June 2023; Additional Remarks: "Visualizing a tile of data for a single time step is taking too much time; Not certain about which time step is better for the task, leading to multiple random navigations."

E4: Motivation: To explore the distribution of global precipitation rates on an annual basis; Outcome: NASA Worldview - GCOM-W1/AMSR2, 2021; Additional Remarks: "It will be good if additional information about data variation is provided; Prefer sequential preview at smaller zoom levels and stepped preview on larger zoom levels."

E5: Motivation: To inspect urban air pollution levels and generate heatmap datasets for subsequent research; Outcome: NASA Worldview - MODIS/Aerosol  Optical Depth, May 2020; Additional Remarks: "Aggregated data summaries will be helpful to determine anomaly and patterns; Frequently go back and forth between time steps due to the absence of contextual information."

}
\end{figure*}

We organize our findings in the \autoref{fig:findings}. We synthesize common opinions from notes and list other insights in the ``additional remarks'' section. According to our study, participants initiate their tasks by identifying target datasets. In this phase, spatial boundary visualizations (E1, E3, E5), preview images and metadata catalogs provide valuable assistance. Upon dataset selection, we find that participants struggle to navigate between time steps due to the absence of contextual guidance (E1-E5). They are\textit{``not certain about which time step to jump after inspecting current one''}(E3) and \textit{``frequently go back and forth between time steps for comparison''}(E5). As a result, most participants (E2-E5) primarily use random navigation between time steps. Only those with explicit intentions, such as investigating specific earth event occurrences (E1, E2), resort to targeted navigation (E1, E2, E5), where they explicitly input dates and times in the temporal selector.

After the initial selection of time ranges, they use sequential navigation (E1, E4) and generate time-lapse videos (E1, E2, E4, E5) to check their selections before downloading. These videos are useful for inspecting the dynamic variation and ensuring the datasets meet their needs (E2). In this phase, some participants (E2, E4) report slow response times of more than 10 seconds due to the large volume of data needed. This long period of waiting time also makes the navigation between time steps difficult (E2, E5). Additionally, some participants (E1, E3) note the absence of more flexible temporal granularity options, with E1 stating that \textit{``solely providing monthly previews are insufficient since the rate of change differs across datasets''}.% Additionally, participants report that \textit{``the spatial comparison feature is useful to find important events''}.

\subsubsection{Salient Time Step Definition}

Our study reveals that the users' priority for time selection varies in different datasets and tasks; a singular definition fails to cater to users' flexible requirements. For instance, for the task of investigating the algal bloom situations (E2), the users expect to locate the time steps that are most severe or degrading with the fastest rate; for the task of exploring the distribution of precipitation rates (E4), the users expect to get a quick overview of the data with minimum frames of download. Therefore, we define \textit{salient time steps} from the following three perspectives:

\paragraph{DF1. Summarizability.} Salient time steps are those that can effectively summarize the \revision{\sidecomment{R1.2c}critical information and trends contained in the dataset. This includes significant alternation in the data's spatial features over time, or rapid changes in data values that signal crucial dynamics.}% These steps capture impactful events or changes, such as major structural changes or rapid variations. % Quantitative Result

\paragraph{DF2. Anomaly.} Salient time steps are those that represent anomalies or outliers in the dataset. These steps reveal the deviation from the norm, such as rare occurrences of unusual precipitation or temperature - structurally different frames from the rest. % Latent Exploration

\paragraph{DF3. Extremum} Salient time steps are those extreme values within a certain spatial or temporal region of the data. These could be maximum or minimum, highlighting key moments of peak activity or low points that require additional attention. % Variation Visualization

% \vspace{3pt}

By proposing these three dimensions, we aim to provide complementary, rather than mutually exclusive multifacet definitions for salient time steps. We intend to integrate these three dimensions in our proposed algorithm in Section \ref{sec:method}, as well as the interactive systems detailed in Section \ref{sec:system}.

\subsubsection{Design Requirements}
\label{sec:design-req}

According to the participants' remarks on interaction, we identified several key design requirements for our system to support users' diverse task and data exploration needs.

\paragraph{DR1. Temporal Context} Users need to be aware of the structure of the dataset to make temporal navigation decisions. The system should provide temporal contexts to facilitate navigation. Here \textit{Temporal Context} refers to the auxiliary information informing users about the current state of selections and temporal patterns of data. This could be the visual clues about the availability and loading state of data, the visualizations for the variation trends, or the similarities between data frames. % markers indicating the anomaly or extremums,

% \paragraph{DR2. Variation \& Similarity} Users needs

\paragraph{DR2. Global \& Regional} The characteristics between the global region, i.e., the entire data-available region of datasets, and regional features, i.e., the users' area of interest differ. Local variations can often be obscured by overarching global changes, hindering users from identifying time steps. Therefore, our system should support the selection of different spatial analysis regions, providing temporal context at varying spatial scales.

\paragraph{DR3. Dynamic \& Static} Time-lapse videos are useful to illustrate the general variation of data, outlining the event dynamics, whereas the specific values of pixels in a static raster provide accurate information. The system should support both dynamic and static visualizations to facilitate user exploration.

\paragraph{DR4. Performance} Long periods of waiting times are detrimental to users' exploration experience, making navigating between time steps cumbersome. The system should provide smooth navigation between time steps. Reasonable waiting time to generate visualizations for 1 frame should be no more than 2 seconds ~\cite{eggerWaitingTimesQuality2012}. This can be achieved through appropriate caching and pre-fetching mechanisms.

These design requirements will guide the design and implementation of the system.

%~\cite{nahStudyTolerableWaiting2004}

\section{Method}
\label{sec:method}

Given a series of temporal data, our goal is to select a subset of time steps that is the most \textit{salient} to users. Intuitively, we aim to evaluate the quality of our current selection iteratively to find the optimal solution. At each iteration, we need to determine if selecting frames $i$ and $j$ yields a better outcome than selecting frames $i$ and $k$. The DP approach is suitable for this setting as it decomposes the problem into multiple non-overlapping sub-problems, solving each one only once, therefore guaranteeing global optimality. % In this section, we first outline our DP approach to select the time steps and elaborate on the cost function in the following subsections.

Formally, let $\mathbf{X} \in \mathbb{R}^{t \times n \times m}$ be a 3D tensor representing the temporal geospatial data, where $t$ is the number of time steps, $n$ and $m$ are the number of rows and columns of the spatial domain, respectively. Let $\mathbf{X}_t$ represent the data at time step $t$. Let the selected salient time steps be denoted as $\mathbf{S} = \{s_1, s_2, \dots, s_k\}$, where $s_i \in [1, t]$ and $s_i < s_{i+1}$. $k$ is the number of salient time steps, specified by users and $k \ll t$.

\subsection{Overview}

We formulate the problem as a DP problem. We initialize a two-dimensional DP table, denoted as $D$, to store intermediate results. Each entry $D(i, j)$ represents the minimum cost associated with frame sequence $\mathbf{S}$ with the number of selected frames $|\mathbf{S}| = i$ and the last selected frame $s_i = j$. Note that the first frame is always selected. The recurrence relation of the DP procedure is defined as follows:

% \begin{equation}
%   D(i, j) = \min_{k \in [j - 1, i]} D(i, k) + D(k, j) + \mathcal{C}(i, j),
% \end{equation}

\begin{equation}
  D(i, j) = \min_{k \in [i - 1, j]} \big\{  D(i - 1, k) + \mathcal{C}(k, j) \big\},
\end{equation}

\noindent where $\mathcal{C}(i, j)$ denotes the cost function, defined as the linear combination of the structural cost $\mathcal{C}_{\mathrm{struc}}$, the statistical variation cost $\mathcal{C}_{\mathrm{stat}}$, and the distance cost $\mathcal{C}_{\mathrm{dis}}$:

\begin{equation}
  \mathcal{C}(i, j) = \alpha \mathcal{C}_{\mathrm{struc}}(i, j) + \beta \mathcal{C}_{\mathrm{stat}} + \mathcal{C}_{\mathrm{dis}}.
\end{equation}

For instance, $D(2, 3)$ denotes the minimum total cost for the time sequence $\mathbf{S} = \{s_1, s_2\}$ with $s_1 = 1$ and $s_2 = 3$. In this specific case, $D(2, 3) = \mathcal{C}(1, 3)$. Additionally, we use a two-dimensional array $p$ to memorize the previous time step $s_{i-1}$ that leads to the minimum cost $D(i, j)$, which is used to backtrack the optimal solution. The overall procedure is summarized in Algorithm~\ref{alg:dp}.

% , which is the sum of the structural cost $\mathcal{C}_{\mathrm{struc}}$, the statistical variation cost $\mathcal{C}_{\mathrm{stat}}$, and the distance cost $\mathcal{C}_{\mathrm{dis}}$. The total cost $\mathcal{C}$ is defined as follows:

% \begin{equation}
%   \mathrm{dp}[i][j] = \min_{k \in [i, j]} \mathrm{dp}[i][k] + \mathrm{dp}[k][j] + \mathcal{C}(i, j)
% \end{equation}

\begin{algorithm}[htbp]
  \caption{Salient Time Steps Selection}
  \label{alg:dp}
  \begin{algorithmic}[1]
  \Require $\mathbf{X}$: Temporal geospatial data with $t$ steps \\
    $k$: Number of time steps to select \\
    $\alpha, \beta$: Parameters for cost function
  %   % $\alpha$, $\beta$, $\sigma$: Parameters for cost function
  \Ensure $\mathbf{S}$: Salient time steps with $|\mathbf{S}| = k$
  \State Initialize $ D \gets \textbf{inf}[1 \ldots t][1 \ldots k],\ p \gets \textbf{zeros}[1\ldots t][1 \ldots k] $
  \State Define cost $\mathcal{C}(i, j) = \alpha \mathcal{C}_{\mathrm{struc}} + \beta \mathcal{C}_{\mathrm{stat}} + \mathcal{C}_{\mathrm{dis}}$
  \For{$i = 1, \ldots, t$}
    \State $D[i, 1] \gets 0$
    \State $D[i, 2] \gets \mathcal{C}(i, 2)$ \Comment{The first frame is always selected}
  \EndFor

  \For{$i = 0, 1, \ldots, t$} \Comment{The ending time step}
      \For{$j = 1, 2, \ldots, \min(i+1, k)$} \Comment{Num. of selected steps}
          \For{$t = j-1, j, \ldots, i-1$}
              \State cost $\gets D[t, j - 1] + \mathcal{C}(t, i)$
              \If{$\text{cost} < D[i, j]$}
                  \State $D[i, j] \gets \text{cost}$
                  \State $p[i, j] \gets t$
              \EndIf
          \EndFor
      \EndFor
  \EndFor

  \State Initialize $\mathbf{S} \gets \textbf{zeros}[1 \ldots k], \ \text{cur} \gets t - 1$
  \For{\( p = k, k-1, \ldots, 1 \)} \Comment{Backtrack to find the solution}
      \State \( \mathbf{S}[p - 1] \gets \text{cur} \)
      \State \( \text{cur} \gets p[\text{cur}, p] \)
  \EndFor
  \State \Return $\mathbf{S}$
  \end{algorithmic}
\end{algorithm}

In our approach, the cost function is critical to the selection result as it quantifies the quality of current selections. \revision{\sidecomment{R3.3\\MR2}When $k$ frames are selected, the data between salient frames is lost, thus compromising the spatial continuity. To minimize information loss, we want each selected time step to be of significant importance and carry as much information as possible.} This requires our cost function to be smaller when the co-existence of two frames in the selection result provides valuable insights, and vice versa. In other words, when two frames are different, the cost decreases. \hide{In this case, we will never choose two identical or quite similar frames as the cost will be large.} Unlike existing methods that measure the difference using information theory ~\cite{zhouKeyTimeSteps2018} or dynamic time wrapping ~\cite{tongSalientTimeSteps2012}, we propose a more expressive and flexible method to define the cost. In the following subsections, we elaborate on the three components of the cost function.

\subsection{Structural Features}

\begin{figure*}[tbhp]
  \centering
  \includegraphics[width=\linewidth]{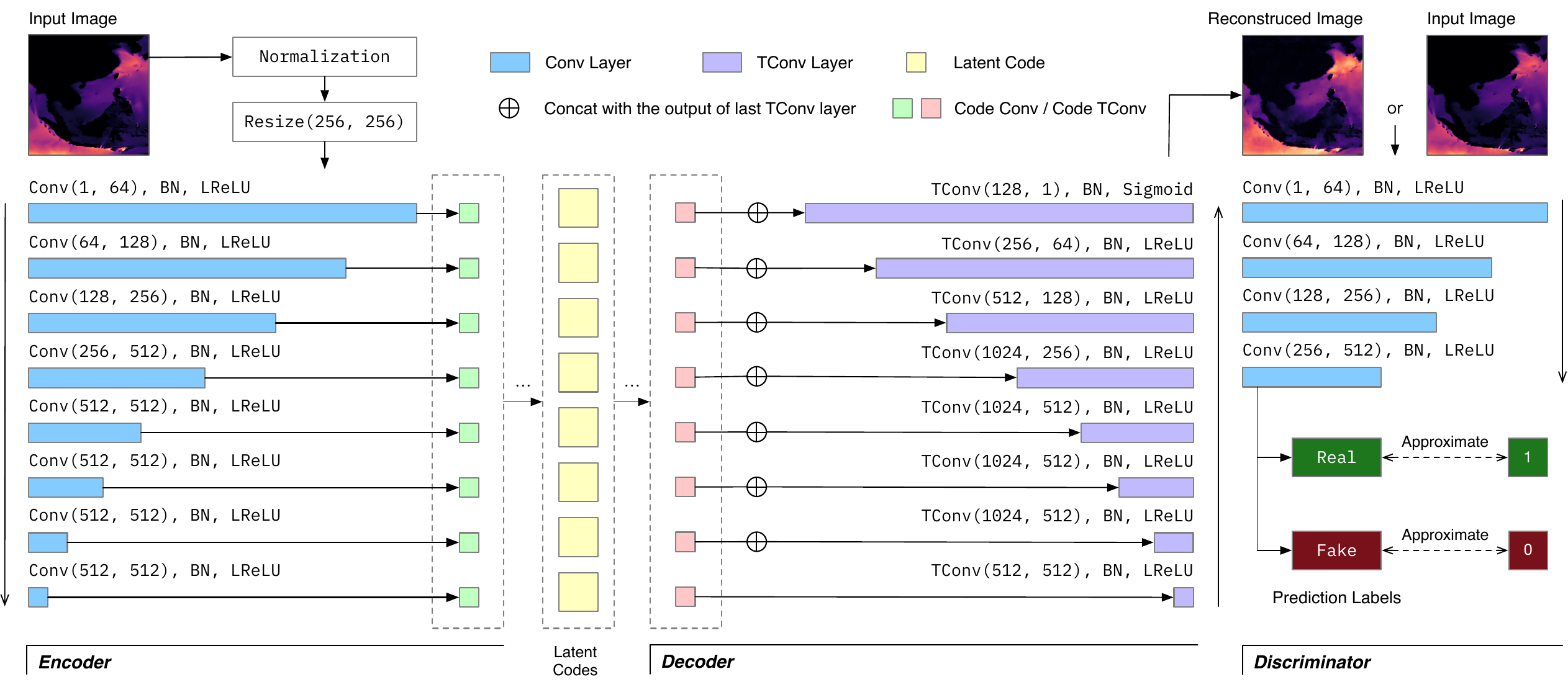}
  \caption{
      The architecture of our network. The encoder operates from top to bottom while the decoder operates from bottom to top. The flow connecting the output of a decoder layer to the next layer's contact operation is omitted for clarity.
  }
  \label{fig:net}
  \Description{
Three sections of network architecture diagrams, from left to right are the Encoder, Decoder, and the Discriminantor. The encoder has eight interconnected convolutional layers, the decoder has eight interconnected deconvolutional layers. The discriminator takes reconstructed images and input images as input, and output prediction labels after four convolutional layers.
  }
\end{figure*}

The structural cost $\mathcal{C}_{\mathrm{struc}}$ quantifies the structural difference between two frames. \revision{\sidecomment{R1.2d}This cost aligns with our definition of time steps DF1 (Summarizability) and DF2 (Anamoly), as it encourages selected frames to reveal the variation of inherent patterns, shapes and spatial characteristics within the data.} Traditional methods often fall short in capturing these variations, especially when data exhibits different semantic spatial relationships but follows similar value distribution. We employ a CNN-based autoencoder to extract these features from data, where the encoder $Enc(\cdot)$ encodes an input frame $\mathbf{X}_t \in \mathbb{R}^{H \times W}$ into a low dimensional latent code $Z_t \in \mathbb{R} ^ {\phi}$. Here $\phi$ is a hyperparameter, denoting the dimension number. The decoder $Dec(\cdot)$ reconstructs the frame $\hat{\mathbf{X}}_t$ from the latent code $Z_t$. The whole procedure can be formulated as: $\mathbf{\hat{X}}_t = Dec ( Enc (\mathbf{X}_t))$. Our goal is to optimize the network parameters to minimize the difference between $\mathbf{X}_t$ and $\hat{\mathbf{X}}_t$. Additionally, we employ a GAN-based discriminator for adversarial training, thereby enhancing the feature extraction ability of our network. \autoref{fig:net} shows the overall architecture of our network.

% \todo[If the network structure should be inspired by / following sth.?] \yhy{I think no.}

\subsubsection{Training Dataset} We collected geospatial data primarily from the NASA Earthdata ~\cite{NASAEarthdata2023} and NOAA Environmental Modeling Center ~\cite{NOAAEnvironmentalModeling}.  The data includes air temperature at different heights, global wave-wind field data, pressure of mean sea level, satellite imagery, etc. Each dataset covers approximately a range of 10 years. For the NaN (Not a Number) values such as land areas for oceanography data, we fill with zeros. \yhydone{Finally, we chose} about 2,000 time steps from each dataset, generating a diverse collection of more than \yhydone{11,000} images as our training dataset.

\subsubsection{Encoder} The encoder \yhydone{takes a single-channel grey-scale image} as input, consisting of 8 convolution layers for multi-scale feature extraction. The feature map size is halved after each convolution operation. Each convolution layer is followed by a batch normalization layer ~\cite{ioffeBatchNormalizationAccelerating2015} and a LeakyReLU activation layer ~\cite{maasRectifierNonlinearitiesImprove}. The output of each convolution layer is \yhydone{fed into} another convolution layer to produce 8 \yhydone{feature vectors}. \yhydone{These vectors are then concatenated as the final latent code: $Z_t = \{z_i\}_1^8$.} \yhydone{Since $Z_t$ is derived from different levels of feature map throughout the network calculation, it contains both high-level and low-level feature information of the original data.}
 % These latent codes are then concatenated to produce the latent code $Z$ for $\mathbf{X}_t$ with size 512.

\subsubsection{Decoder} The decoder generates the reconstructed frame $\hat{X}$ from the latent code $Z_t$. \yhydone{Each sub-feature vector $z_i$ of $Z_t$ is first fed into 8 deconvolution layers separately to produce 8 feature maps.} \yhydone{} \yhydone{Then, these feature maps go through a series of deconvolution layers for upsampling.} \yhydone{Formally,} the first layer takes $z_1$ as input. Afterward, the output of the $i$-th deconvolution layer is concatenated with the feature map of $z_i$ and fed into the $(i+1)$-th deconvolution layer. Each layer is followed by a batch normalization and LeakyReLU except the last layer \yhydone{which uses a Sigmoid activation} to ensure the output values are in the range of $[0, 1]$.

The output of the decoder's last layer is the reconstructed frame $\hat{X}$. We \yhydone{calculate the reconstruction loss with the mean squared error (MSE) to guide the network training:}

\begin{equation}
  \mathcal{L}_{rec} = \frac{1}{H \times W} \sum_{i=1}^{H} \sum_{j=1}^{W} \left\| \mathbf{X}_t(i, j) - \hat{\mathbf{X}}_t(i, j) \right\| _2,
\end{equation}
\yhydone{where $(H, W)$ denotes the image size.}

\subsubsection{Discriminator} The discriminator network \yhydone{$Dis(\cdot)$} serves as an additional supervision to the reconstruction quality. \yhydone{We adapt the design of GANs~\cite{Goodfellow2014GenerativeAN} to perform adversial training. The discriminator takes \yhydone{$\mathbf{X_t}$ and $\hat{\mathbf{X}}_t$} as input and produces two feature maps indicating the network prediction label. We train $Dis(\cdot)$ to discriminate the reconstructed image thus encouraging the autoencoder to generate a reconstructed image that cannot be distinguished from the original one. Formally, we optimize the network by minimizing:}

\begin{equation}
  \mathcal{L}_{dis} = \left\| Dis(\mathbf{X_t}) - \mathbf{I}_{r} \vphantom{Dis(\hat{\mathbf{X}_t})} \right\| _2 + \left\| Dis(\hat{\mathbf{X}_t}) - \mathbf{I}_{f} \right\| _2,
\end{equation}
\yhydone{where $\mathbf{I}_{r}$ and $\mathbf{I}_{f}$ are the ground truth label of real or fake image.}

\subsubsection{Structural Cost}

% \yhy{\sout{We first concatenate 8 latent codes $\{ z_i \}_1^8$ of the encoder to produce the latent code $Z_t$ with the size of 512.}} \yhy{We measure} the structural similarities of frames $i$ and $j$ by \yhy{calculating the cosine similarities of their latent codes}:

We measure the structural similarities of frames $i$ and $j$ by calculating the cosine similarities of their latent codes:

\begin{equation}
  S(Z_i, Z_j) = \frac{Z_i \cdot Z_j}{\|Z_i\| \cdot \|Z_j\|}.
\end{equation}

% \yhy{\sout{where $Z_i$ and $Z_j$ are the latent codes of frames $i$ and $j$.}}
\noindent Then, the structural cost is then defined as:

\begin{equation}
  \mathcal{C}_{\mathrm{struc}}(i, j) = \frac{1}{1 + e^{-5 \big[ S(Z_i, Z_j) - 0.5 \big] }}.
\end{equation}

The sigmoid-like function maps the cosine similarities to the range of $[0, 1]$, which is shifted and scaled to enlarge the difference between similar frames, as we discover that the cosine similarities between the frames are usually within the range $[0.5, 1.0]$. The cost is minimized when the two frames are structurally similar.

\subsubsection{Training}\label{sec:training} All training images are first normalized and resized to the network input size of $256 \times 256$. \yhydone{Our model is implmented with PyTorch ~\cite{paszke2019pytorch}. For model training, we use the Adam optimizer ~\cite{Kingma2014AdamAM} with an initial learning rate of $10^{-4}$ and a batch size of 128. The network converges after 350 epochs. The whole training process is conducted on a Linux server with a 64-core Intel Xeon CPU, an NVIDIA GeForce 3090 GPU, and 128 GB of memory.}

% Due to the diverse geospatial data distribution in the training data and the inherent generalizability of the encoder-decoder network, our network can be used to encode various types of geospatial data and generate meaningful latent codes for structural cost calculation.

% \yhy{\sout{Due to the diverse geospatial data distribution in the training data and the inherent generalizability of the encoder-decoder network, our network can be used to encode various types of geospatial data and generate meaningful latent codes for structural cost calculation.}}

% \textcolor{red}{<- not training details.}

\subsection{Statistical Variation}

The statistical variation cost $\mathcal{C}_{\mathrm{stat}}$ measures the variation of aggregated data\revision{, assisting users in identifying anomalies or obtaining a statistical summary of the dataset.} We consider three aggregation operations: \texttt{MAX}, \texttt{MIN} and \texttt{AVG}. These aggregation operations are commonly used in geospatial analysis, as suggested in our need-finding study. The aggregation can be applied either globally on the entire spatial region of $\mathbf{X}$, or regionally, focusing on user-specified regions. \revision{\sidecomment{R1.2d}This cost aligns with our definition DF2 (Anomaly) and DF3 (Extremum), as it encourages the selection of frames representing local extrema when using \texttt{MAX} or \texttt{MIN} aggregations, or anomalies when using \texttt{AVG} aggregation, especially when applied to a local region. % This is useful when users are interested in identifying anomalies in the dataset or obtaining a statistical summary of the dataset.
} Let $\textbf{v} = \{ v_i \}_1^t$ denote the aggregated value at time step $i$. We first normalize the aggregated value:

\begin{equation}
  \hat{v}_i = \frac{v_i - \mathrm{nanmin}_t(\mathbf{v})}{\mathrm{nanmax}_t(\mathbf{v}) - \mathrm{nanmin}_t(\mathbf{v})},
\end{equation}

\noindent where $\mathrm{nanmax}$ and $\mathrm{nanmin}$ indicate the maximum and minimum values of $\mathbf{v}$, excluding NaN values. We then define the statistical variation cost as follows:

\begin{equation}
  \mathcal{C}_{\mathrm{stat}}(i, j) = -\tanh(|\hat{v}_i - \hat{v}_j|) + 1.
\end{equation}

$\mathcal{C}_{\mathrm{stat}}$ is designed to be minimized when the aggregated value of two frames is substantially different, thereby encouraging the selection of statistically different frames. This is helpful to identify the local extremum and anomalies of the dataset, as detailed in Section \ref{sec:exp_anamoly}. The $\tanh$ function ensures that the range of cost lies between 0 and 1, in alignment with $\mathcal{C}_\mathrm{struc}$. When selecting frames, users can adjust $\alpha$, $\beta$ and the aggregation method to control the priorities for frame selection, where $\alpha + \beta = 1$.

\begin{figure}[htbp]
  \centering
  \includegraphics[width=\singlecolumnwidth]{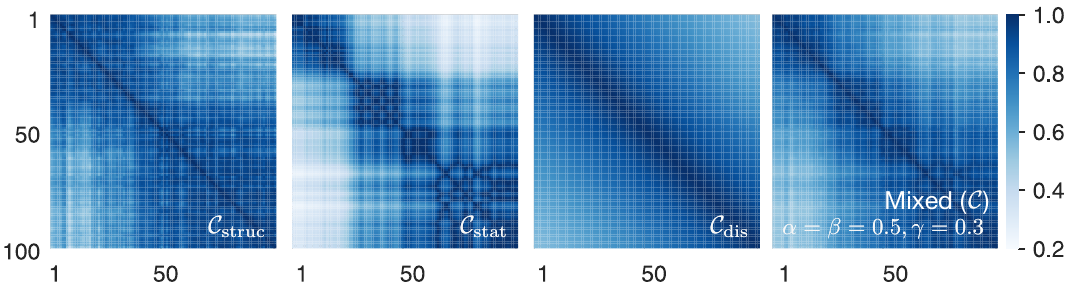}
  \caption{
    The cost matrix for the first 100 frames from data \texttt{tmp2m} (Air Temperature at 2 Meters), using \texttt{AVG} as aggregation method for $\mathcal{C}_{\mathrm{stat}}$. Each item $(i, j)$ in the matrix denotes the cost value for frames $i$ and $j$.
  }
  \label{fig:matrix}
  \Description{
Four colorized matrices show the cost matrix for structural cost, statistical variation cost, distance cost, and mixed cost. Each matrix is a 100 by 100 matrix, with each item denoting the cost value for frames $i$ and $j$. These matrices exhibit different patterns, with the structural matrix showing a diagonal line with multiple vertical lines with low values; the statistical variation cost matrix showing more lines with low values, showing a more pronounced change trend; the distance cost matrix showing a line with a slope, and the mixed cost matrix showing a combination of the three patterns.
}
\end{figure}

\autoref{fig:matrix} shows an example of cost matrix for $\mathcal{C}_{\mathrm{struc}}$ and $\mathcal{C}_{\mathrm{stat}}$, along with the mixed cost matrix with $\mathcal{C}_{\mathrm{dis}}$. It can be observed that the statistical variation, which is the average temperature in this case, exhibits significant changes ahead of the structural features, which emcompasses the spatial features of the data. The combination of the two costs allows users to achieve more comprehensive selection results.

\subsection{Distance Penalty}

\begin{figure}[htbp]
  \centering
  \includegraphics[width=\singlecolumnwidth]{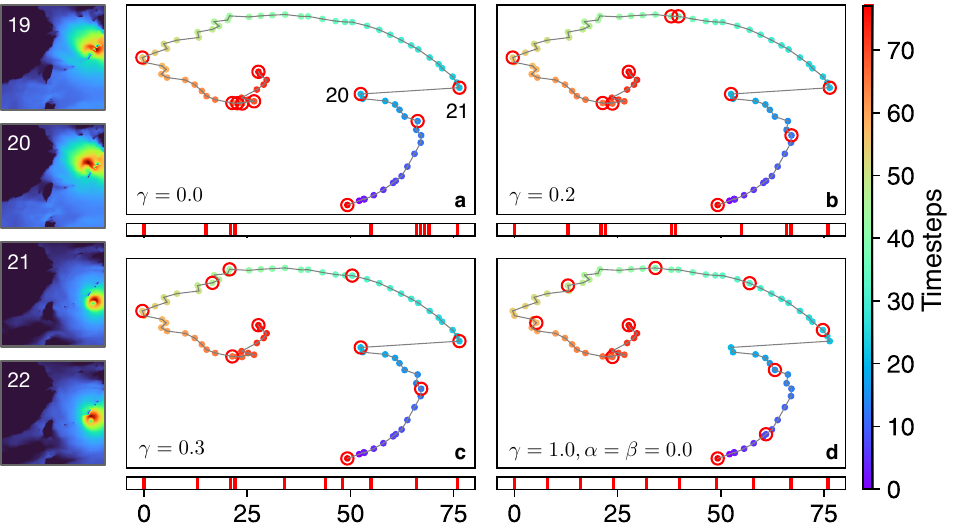}
  \caption{
    The impact of the distance cost parameter \(\gamma\) on the selection of salient time steps. Subfigures a, b, c, and d show the results with different $\gamma$s. Each dot represents a frame's latent code $Z_i$, dimension-reduced with t-SNE ~\cite{maatenVisualizingDataUsing2008} and colored in chronological order. Salient time steps are marked in red circles. On the left are the colorized frames 19 to 22, where drastic changes occur between frames 20 to 21, as reflected in the latent space. The selection is performed on 80 frames of \texttt{hs} (significant height of wind and swell waves) data of the East China Sea area with k = 10.
  }
  \label{fig:cdis}
  \Description{
Four subfigures, each showing the data's trajectories in the latent space. When $\gamma$ is small, the selection tends to cluster in a short period.
  }
\end{figure}

The distance cost $\mathcal{C}_{\mathrm{dis}}$ is introduced to ensure that the selected frames are well distributed across the temporal domain. The distance cost is defined as follows:

\begin{equation}
  \mathcal{C}_{\mathrm{dis}}(i, j) = -\gamma \tanh(\frac{|i - j|}{\sigma n / k}) + 1,
\end{equation}

\noindent where $|i - j|$ is the distance between two frames, $n$ is the total number of frames, $k$ is the number of frames to select, and $\sigma$ and $\gamma$ are two hyperparameters. Specifically, $\gamma$ controls the weight of the distance cost, and $\sigma$ controls the rate of decay as the distance increases. In our experiments, we set $\gamma = 0.3$ and $\sigma = 1.0$ to add a slight penalty to the distance between two frames. The distance cost acts as a regulatory term that penalizes the close clustering of selected frames, thereby encouraging a more balanced and informative selection. When solely using the distance cost ($\gamma = 1, \alpha = \beta = 0$), evenly distributed time steps will be selected (\autoref{fig:cdis}.d).

The distance cost is essential for preventing the unintended clustering of time steps, a potential drawback of the global optimality in DP. While DP aims to find the optimal solution, it may select consecutive frames when high variation occurs in a short period, as shown in \autoref{fig:cdis}.a. This can obscure important temporal patterns. The distance cost mitigates this problem by penalizing consecutive frames, encouraging the selection of more evenly distributed time steps, while still selecting frames with significant variations. As shown in \autoref{fig:cdis}.c and d, the distance cost prevented the unintended clustering of frames 66 - 68, while still selecting frames 20 and 21 as salient time steps due to the substantial changes between them.

% \todo[A figure illustrating with or without distance cost]

% \todo[A figure with gridded matrix and selection result for global-hs]
% \subsubsection{Periods}

\section{System Design}
\label{sec:system}

To facilitate interactive exploration of geospatial data and temporal selection, we design and implement a web-based system that integrates our approach with context visualizations. The design requirements from our need-finding study are carefully considered, enabling efficient temporal exploration for geospatial data. % In this section, we introduce the user interface and the implementation details of our system.
% \subsection{Architecture}

\subsection{User Interface}

\autoref{fig:teaser} displays the user interface of our system. For clarity, we use bold letters to refer to the labels in \autoref{fig:teaser} unless otherwise specified throughout this section.

\subsubsection{Map View}

The map view displays heat maps for geospatial data transformed through color maps, providing users with insights about the current data frame. On mouse hover, the data value corresponding to the coordinates is shown. The frames' global max, min and average values are displayed at the bottom of the Control View (\bs{c}). Users can select a specific spatial boundary to enter the regional mode for analysis (\bs{g}), in line with Design Requirement 2 (DR2). When in regional mode, the salient frame calculations and context visualizations are performed within the selected region.

\subsubsection{Timeline View}

The Timeline View serves as the cornerstone of our system, offering temporal selection and exploration functionalities. At its center is the time series of the dataset. Users first need to drag across a span of time to select the \textit{focus range}, representing the time range they are currently interested in. The selection of salient times will be executed within the range. This design stems from our observation that users typically don't aim to explore the whole available range of the dataset.

\begin{figure}[htbp]
    \centering
    \sidecomment{R1.3\\MR7}
    \revisionbox{
        \includegraphics[width=\singlecolumnwidth]{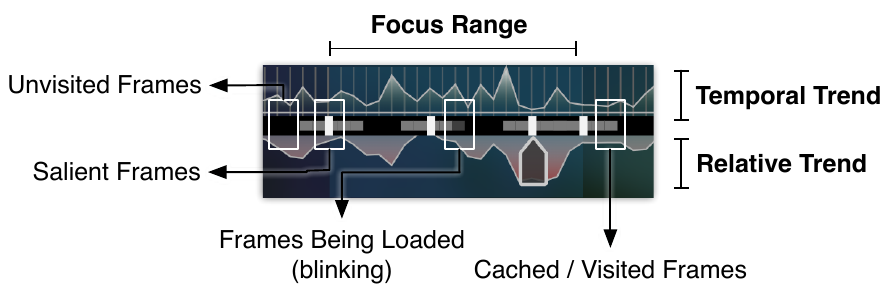}
    }
    \caption{
        Visual encodings for the frame status.
    }
    \label{fig:timeline}
    \Description{
A diagram showing the visual encodings of the frame status. The background is a screenshot of the timeline, with arrows pointing to different components of the UI.
    }
\end{figure}

Different visual encodings are employed to represent the data frame status, as illustrated in \autoref{fig:timeline}. Salient frames are marked in white, while frames visited by users or preloaded are marked in gray. A blinking mark indicates a frame that's currently being loaded. Given that we preload frames adjacent to the salient frames, these loading indicators can help inform users of the loading progress.

To address DR1, we incorporate 2 context visualizations on the timeline. The line chart on the top (\bs{d}) represents the \textit{Temporal Trend}, indicating the data's temporal variations. The line chart on the bottom (\bs{e}) represents the \textit{Relative Trend}, indicating the differences of other frames relative to the current frame, updated every time the user navigates to a new frame. Both context visualizations can be configured to display the structural variations or three types of statistical variations.

\begin{figure}[htbp]
    \centering
    \includegraphics[width=\singlecolumnwidth]{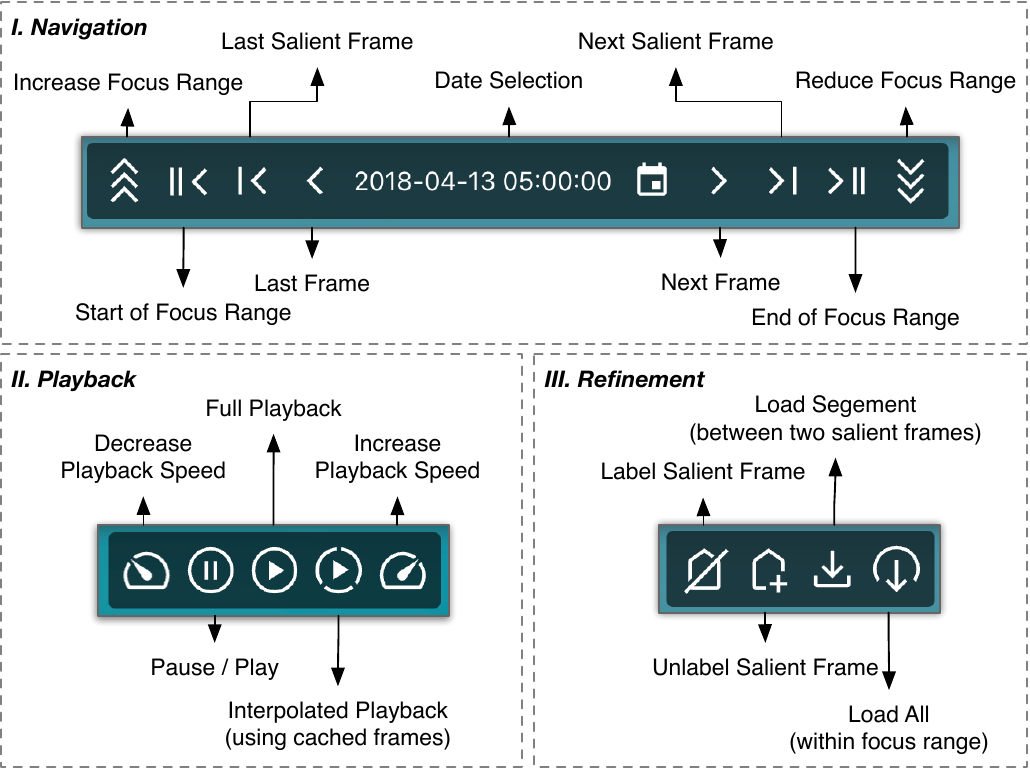}
    \caption{
        Functions provided by the Timeline View.
    }
    \label{fig:functions}
    \Description{
        Three subfigures, each showing one section of function panels below the Timeline View, with arrows pointing to different components of the UI.
        The first panel is the Navigation panel, the functions from left to right are: Increase Focus Range; Start of Focus Range; Last Salient Frame; Last Frame; Date Selection; Next Salient Frame; End of Focus Range; Reduce Focus Range;
        The second panel is the Playback panel, the functions from left to right are: Decrease Playback Speed, Pause / Play; Interpolated Playback (using cached frames); Increase Playback Speed;
        The third panel is the Refinement panel, the functions from left to right are: Label Salient Frame; Unlabel Salient Frame; Load Segment (between two salient frames); Load All (within focus range)
    }
  \end{figure}

Three sections of the function panel positioned below the Timeline View (\bs{f}) provide a comprehensive set of auxiliary tools to assist temporal selection. \autoref{fig:functions} lists their functionalities. The Navigation section \revision{(\autoref{fig:functions}.\bs{I})} assists users in navigating between time steps, while the Playback section \revision{(\autoref{fig:functions}.\bs{II})} facilitates the creation of time-lapse videos on the map. The interpolated playback utilizes locally cached frames, creating data playback by smoothly transitioning between these frames. The full playback requires full data to be downloaded, potentially incurring significant waiting time. \revision{\sidecomment{R2.1\\MR8}The Refinement section (\autoref{fig:functions}.\bs{III}) allows users to manually label or unlabel salient frames. When performing salient frame selections, frames manually labeled by the user are guaranteed to be selected, while unlabeled frames will be excluded from the selection result.} Both user-labeled and system-generated salient frames are displayed above the timeline for reference (\bs{i}).

\subsubsection{Control View}

The Control View allows users to select the dataset and specify the parameters $\alpha$, $\beta$, and $k$ for salient frame selection (\bs{a}). A circular indicator indicates the current parameter states. The visualization of the dimension-reduced latent space (\bs{b}) \revision{ within the focus range} provides additional structural variation information for the data, assisting users in identifying crucial moments of data variation. This visualization is the same as we provided in \autoref{fig:cdis}. \revision{\sidecomment{R2.4\\MR10}To reduce visual clutter, we constrain the maximum number of points visualized in the latent space to 500. If the number of frames in the focus range exceeds 500, we sample time steps in the latent space evenly. All salient time steps will be displayed, irrespective of the constraint.
}Salient frames are highlighted here with red circles, aligning with the information on the timeline. Users can navigate to specific frames by clicking on a frame in the latent space. Additionally, users can adjust the colormap and mapping functions at the bottom (\bs{c}).

\subsubsection{Playback View}

The Playback View (\bs{h}) offers additional dynamic information for the data, addressing DR3. It utilizes locally cached data frames to cyclically playback the data within the focus range, allowing users to simultaneously inspect both static and dynamic information. This view is independent of the previously mentioned playback functionalities on the map.

% \subsubsection{Similarity Matching}
% \subsection{Variation Visualization}
% \subsection{User-driven Selection Refinement}
\subsection{Implementation}

Our system is developed using modern web technologies including React.js, Webpack, and TypeScript. We use D3 \cite{BostockD3DataDriven2011} for data visualization and Mapbox GL JS for map rendering. We employ the GPU-accelerated t-SNE ~\cite{chanGPUAcceleratedTdistributed2019} to accelerate visualization for latent codes. The system is deployed on the same server as described in Section \ref{sec:training}. Addressing DR4, we minimize the computation cost by caching the latent code for different spatial queries and employing paralleled aggregation computation for better performance.

\section{Case Study}

To validate the efficacy of our system, we apply it to two geospatial datasets: RDI ~\cite{shenSimpleMethodsSatellite2019} (Red Tide Detection Index) for algal bloom detection and Cloud Moisture Imagery ~\cite{GOES16BandReference2016} for hurricane analysis. The user study was conducted with Expert 2 (E2) and Expert 5 (E5) from our need-finding study, after which we collected user feedback and comments.

% \subsection{Global Wave Hindcast}

% Points: Dataset, Expoert, Interaction, Design Requirements, Outcome.

\subsection{Algal Bloom}

The first case study was conducted with E2. Algal blooms, especially harmful ones (HABs) are one of the major threats to the marine ecosystem and human health. Early detection and warning of HABs is crucial for environmental management and public health. RDI is introduced to monitor HABs from ocean color satellite imageries, employing a three-band blended reflectance model that uses data from MERIS, MODIS and GOCI ~\cite{shenSimpleMethodsSatellite2019}. The dataset we used is a 3-year time series that covers the Yangtze River Estuary in China, with a spatial resolution of 500 m and a temporal resolution of 5 - 10 days. The dimension of the data is $486 \times 4152 \times 2642$.

\begin{figure}[htbp]
    \centering
    \sidecomment{R1.3\\MR7}
    \revisionbox{
    \includegraphics[width=\singlecolumnwidth]{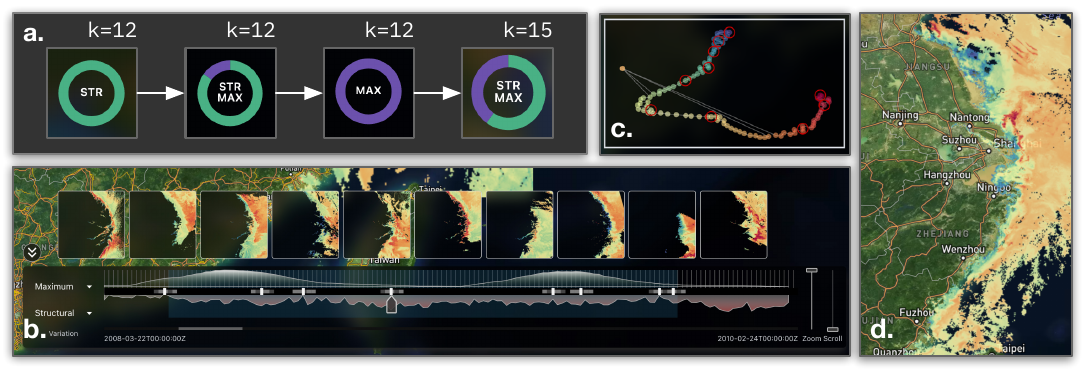}
    }
    \caption{
        User interaction and results involved in the case study for RDI data. The user's choices for selection parameters are enumerated in (a), with the extent of the green arc in the hollow circle representing the value of $\alpha$.
    }
    \label{fig:case-rdi}
    \Description{
Four subfigures. (a) lists the users parameter choices: $\alpha = 1.0, \beta = 0.0, k = 12$; $\alpha = 0.8, \beta = 0.2, k = 12$; $\alpha = 0.0, \beta = 1.0, k = 12$; $\alpha = 0.6, \beta = 0.4, k = 15$, using \texttt{MAX} aggregation. (b) shows a screenshot of the timeline view with selections of RDI data. The temporal trend shows a cyclical pattern of RDI values. (c) displays the latent space for the current selection. (d) shows the recorded algal bloom event in May 2009.
    }
  \end{figure}

Initially, the user employed a smaller zoom level and quickly selected several focus ranges on the timeline using the default number of time steps $k=12$. The salient frames were generated almost instantly (DR4). Within the following seconds, the salient frames and their adjacent frames were sequentially downloaded. Since a higher RDI value indicates a more severe algal bloom situation, the user added the weight of statistical variation $\beta$ to 0.2 and set the aggregation to \texttt{MAX}, resulting in a different set of salient frames. Then, the user tried setting $\beta$ to 1, obtaining the frames that show the peaks and troughs of RDI maximum values. Subsequently, the user reset $\alpha = 0.6, \beta = 0.4$ and increased $k$ to 15 to get a more balanced selection (\autoref{fig:case-rdi}.a), navigating between the time steps to obtain data overview.

Then, the user selected \texttt{MAX} on the Temporal Trend above the timeline and observed an annual cyclical trend in the RDI values (\autoref{fig:case-rdi}.b): Algal blooms primarily occur between April and August each year, peaking around June. The user navigated to one of the peak Frames 156 (\autoref{fig:case-rdi}.d), which is a recorded algal bloom event in May 2009. The user then selected \textit{Structural} in the Relative Trend to compare the spatial characteristics of this event with other frames. Afterward, the user zooms in and examines the data around this period. The latent space within this range is shown in \autoref{fig:case-rdi}.c. By observing the patterns in the latent space, the user further navigated data frames at turning points for observation. Eventually, the user chooses to request the full data for frames 950 - 1200.

The user remarked that our system offers a flexible approach to temporal navigation. With the swift response speed of salient time selection, the user experience enhanced significantly compared with existing data access portals. The user also appreciated the value of contextual visualizations in making informed navigation between timesteps and identifying important events (DR1), and expressed the ``\textit{desire to integrate this approach into the workflow of existing systems}''.

\subsection{Hurricane}

In our second case study with E5, we focus on Hurricane Dorain, a category 5 hurricane that struck the Bahamas and the Southeastern US in 2019. The data was produced by the Advanced Baseline Imager (ABI) instrument from the GOES-16 satellite. We use the Cloud Moisture Imagery Product (CMIP), which is a 16-band dataset that covers the entire Western Hemisphere. We use the Continental US area, Band 13 of the product, which is the Clean Infrared Window, used to monitor clouds, storm systems, and hurricane intensity ~\cite{GOES16BandReference2016}. We populate data from August to September 2019 in a 20-minute temporal resolution. The dimension of the data is $1653 \times 994 \times 1532$.

\begin{figure}[htbp]
  \centering
  \includegraphics[width=\singlecolumnwidth]{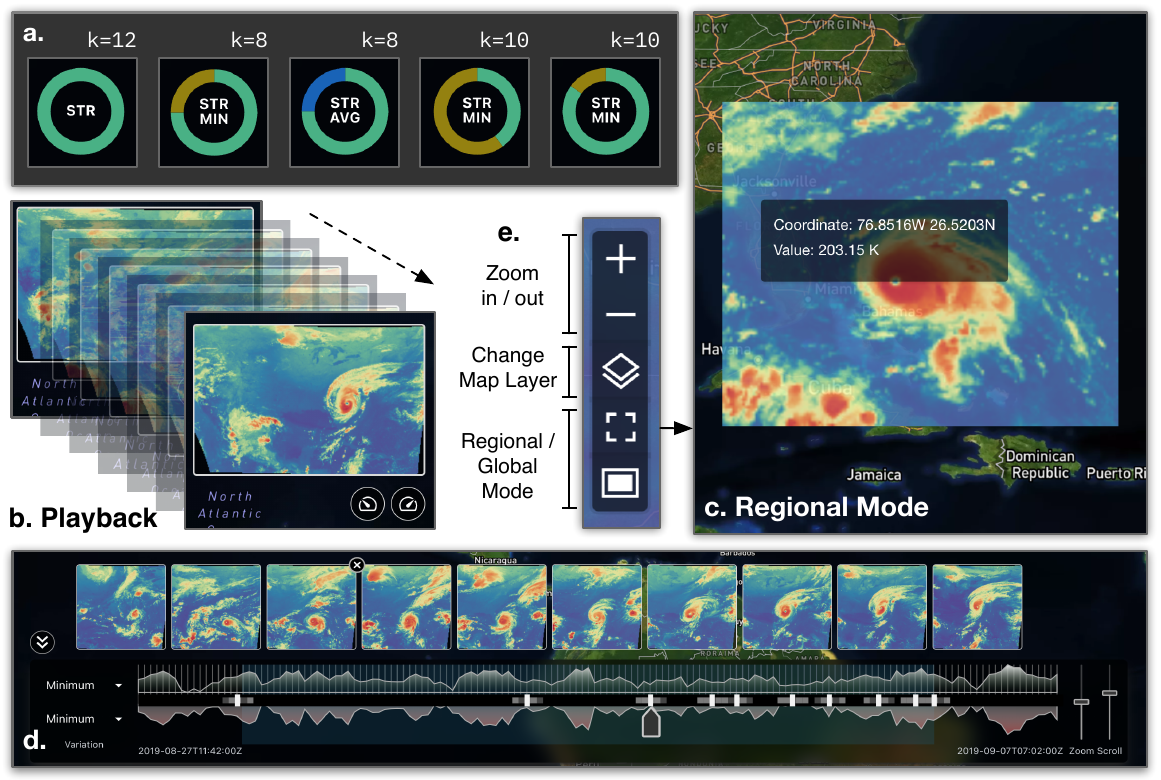}
  \caption{
      User interaction and results involved in the case study for CMIP data. (d) shows the selection results between August 28 to September 6, with $\alpha = 0.8, \beta=0.2$, using \texttt{MIN} aggregation. (e) describes the functions provided by the toolbar on the map's top-left corner.
  }
  \label{fig:case-cmipc}
  \Description{
    Five subfigures. (a) lists the user's parameter choices. $\alpha = 1.0, \beta = 0.0, k = 12$; $\alpha = 0.75, \beta = 0.25, k = 8$; $\alpha = 0.75, \beta = 0.25, k = 8$; $\alpha = 0.4, \beta = 0.6, k = 10$; $\alpha = 0.8, \beta = 0.2, k = 10$, using \texttt{MIN} aggregation except the third query, using \texttt{AVG} aggregation. (b) illustrates the playback process of the interpolated playback view through stacking multiple screenshots. (c) displays the hurricane eye in the regional mode. (d) is a screenshot of the selection result of CMIPC data. (e) illustrates the functions provided by the toolbar on the top left corner of the map. The functions from top to left: Zoom in / out; Change Map Layer; Regional / Global Mode.
  }
\end{figure}

The imagery value represents the Top-Of-Atmosphere (TOA) brightness temperature, which drastically decreases in hurricanes and storms due to their deep convection. The user first selected \texttt{MIN} in the temporal trend to view overall brightness temperature changes over the two months, then quickly navigated to the tropical wave on August 24 and the formation of the hurricane on August 28. The user then selected a focus range, finding salient frames in between to discern the evolution of the hurricane eye. \autoref{fig:case-cmipc}.d displays this selection result. The data variations during the early stages of tropical waves are relatively stable, leading to the selection of only two salient frames. As the hurricane rapidly expands and moves northward after its formation, the selections in the latter half of the focus range become much denser.

The user then observed the interpolated preview in the upper right corner, which dynamically displays the hurricane activity (\autoref{fig:case-cmipc}.b). Initially, with fewer locally cached frames, the playback relies heavily on the interpolated values, transitioning between cached frames. As the user navigated the timeline and more pre-loaded frames became available, the preview became smoother, revealing the data's original dynamics (DR3). Loading full data for playback took roughly 40 seconds, but the playback was available immediately, progressively loaded and continuously increasing fidelity, eliminating wait times (DR4).

Subsequently, the user selected the spatial boundary above the Bahamas to inspect the hurricane eye and observed that on September 1, at 5 pm, the brightness temperature near the hurricane eye had dropped to 203 Kelvin (\autoref{fig:case-cmipc}.c). The user zoomed in on the timeline for a granular view of the hurricane's development within the spatial boundary and navigated to other similar frames with the help of relative trend visualizations (DR2), eventually requesting the download of frames from 800 to 1000.

The user pointed out that our interpolated playback mechanism reduces the waiting time and greatly enhances the user experience for creating time-lapse videos. The diverse contextual visualizations, combined with the capability to select spatial boundaries help users in making more informed temporal navigations and identifying important regional events.

\section{Evaluation}

\revision{

In this section, we first conduct quantitative experiments on diverse geospatial datasets, thereby demonstrating our approach's ability to identify salient characterized by summarizability, anamoly and extremum.
% Then, we present the summarizability and performance of our method.
Then, we conduct expert interviews to assess the usability of our system compared with existing systems mentioned in Section \ref{sec:needfinding}.

}

\subsection{Datasets and Metrics}

\subsubsection{Datasets}
Five datasets are collected for evaluation. Temperature at 2 Meters (\texttt{tmp2m}) and Air Pressure at Mean Sea Level (\texttt{prmsl}) are generated from the Climate Data Assimilation System ~\cite{yangClimateDataAssimilation}, an observational information reconstructed from historical and present states of earth climate. Significant Wave Height (\texttt{hs}) and Mean Wave Length (\texttt{lm}) are extracted from wave hindcast data from WAVEWATCH III ~\cite{tolmanUserManualSystem2014} model, where \texttt{lm} data covers East China Sea area and \texttt{hs} covers global range. Cloud and Moisture Imagery Product (\texttt{cmipc}) ~\cite{GOES16BandReference2016} is the same dataset used in our hurricane case study. \autoref{tab:dataset} lists the dimensions, temporal and spatial coverage of the datasets. During the experiment, we are not using the entire time span but using a subset of timesteps as the \textit{focus range}, in alignment with our interactive system.

\begin{table*}[tbhp]
  \caption{Datasets used for evaluation. Temporal resolution (Res.) indicates the time interval between two consecutive time steps, and coverage (Cov.) indicates the time span of the dataset.}
  \small
  \renewcommand{\arraystretch}{0.8}
  \label{tab:dataset}
  \begin{tabular}{llllll}
    \toprule
    Code & Description & Source & Spatial Region & Dimension ($t \times n \times m$) & Temporal Res. / Cov. \\
    \midrule
    \texttt{tmp2m} & Air Temperature at 2 Meters & CDAS-1 ~\cite{yangClimateDataAssimilation} & Global & 1588 $\times$ 880 $\times$ 1760 & 1 d / 4 years \\
    \texttt{hs} & Significant Height of Wind and Swell Waves & WAVEWATCH III ~\cite{tolmanUserManualSystem2014} & Global & 4632 $\times$ 311 $\times$ 720 & 3 h / 2 Years \\
    \texttt{prmsl} & Air Pressure at Mean Sea Level & CDAS-1 ~\cite{yangClimateDataAssimilation} & Global & 1463 $\times$ 361 $\times$ 720 & 6 h / 1 Year \\
    \texttt{cmipc} & Cloud Moisture Imagery, Band 13 (10.3 $\mu m$) & GOES-16 ~\cite{GOES16BandReference2016} & Continental US & 1653 $\times$ 994 $\times$ 1532 & 20 min / 20 Day \\
    \texttt{lm} & Mean Wave Length & WAVEWATCH III ~\cite{tolmanUserManualSystem2014} & East China Sea & 234 $\times$ 551 $\times$ 401 & 3 h / 1 Month \\
    % \texttt{rdi} & Red Tide Detection Index ~\cite{shenSimpleMethodsSatellite2019} & Sentinel-3A/B & Yangtze River Estuary & 38 $\times$ 4152 $\times$ 2642 & 1 - 10 d / 1 Year \\
    \bottomrule
  \end{tabular}
\end{table*}

\subsubsection{Method}

For each dataset, we conduct two phases of experiments. \revision{The first phase evaluates the summarizability of the selected salient time steps compared with even selection and arc-based selection \cite{porterDeepLearningApproach2019}. Even selection sample $k$ time steps uniformly across the whole time span. Arc-based selection selects time steps from the t-SNE dimension reduced latent space via path simplification with Euclidean distance and accumulated angle threshold.} For this experiment, we only consider the structural cost by setting $\alpha = 1$ and $\beta = 0$, as the $\mathcal{C}_{\mathrm{struc}}$ aims to maximize the structural difference between the frames. We quantify the summarizability of selection results by comparing the piece-wise linear interpolated data with the original data. Let selected time steps be $\mathbf{S} =  \{s_i\}_i^t$ and original data be $\mathbf{X}$, the piece-wise linear interpolated data $\mathbf{\tilde{X}}$ is defined as:

\begin{equation}
  \mathbf{\tilde{X}}t =
  \begin{cases}
  \mathbf{X}_t & \text{if } t \in \mathbf{S} \\
  \mathbf{X}_{s_i} + \frac{t - s_i}{s_{i+1} - s_i}(\mathbf{X}_{s_{i+1}} - \mathbf{X}_{s_i}) & \text{otherwise} \\

  \end{cases}
\end{equation}

Following Porter et al. ~\cite{porterDeepLearningApproach2019}, we use PSNR (in dB) and RMSE to measure the quality of the interpolation. \revision{\sidecomment{R3.2\\MR1}Starting with $k = 5$, we select different $k$ values and compare the interpolated data against the results obtained using even selection. In alignment with practical applications, $k$ does not exceed 30\% of the total time steps. Since arc-based selection employs different thresholds to generate salient frames, the number of time steps $k$ is unpredictable. Therefore, we apply a specific threshold (distance threshold $\epsilon = 0.3$, angle threshold $\theta = \pi / 4$, mixing factor $\alpha = 0.5$) to get $k'$ frames and use the same $k'$ with our method and even selection for comparison.
}

The second phase focuses on the impact of statistical variation cost $\mathcal{C}_{\mathrm{stat}}$. We set $k$ as a fixed value $k'$, and progressively adjust $\beta$, the weight of $\mathcal{C}_{\mathrm{stat}}$ from 0 to 1. The results of different $\beta$ are displayed, alongside the aggregated data value at each time step. % illustrating the transition of time step selection -- from prioritizing structural differences to statistical differences.

\subsubsection{Meaning of the Plots}
\label{sec:plot_meaning}

\begin{figure*}[tp]
  \centering
  \sidecomment{R2.5\\MR7}
  \revisionbox{
    \includegraphics[width=0.98\linewidth]{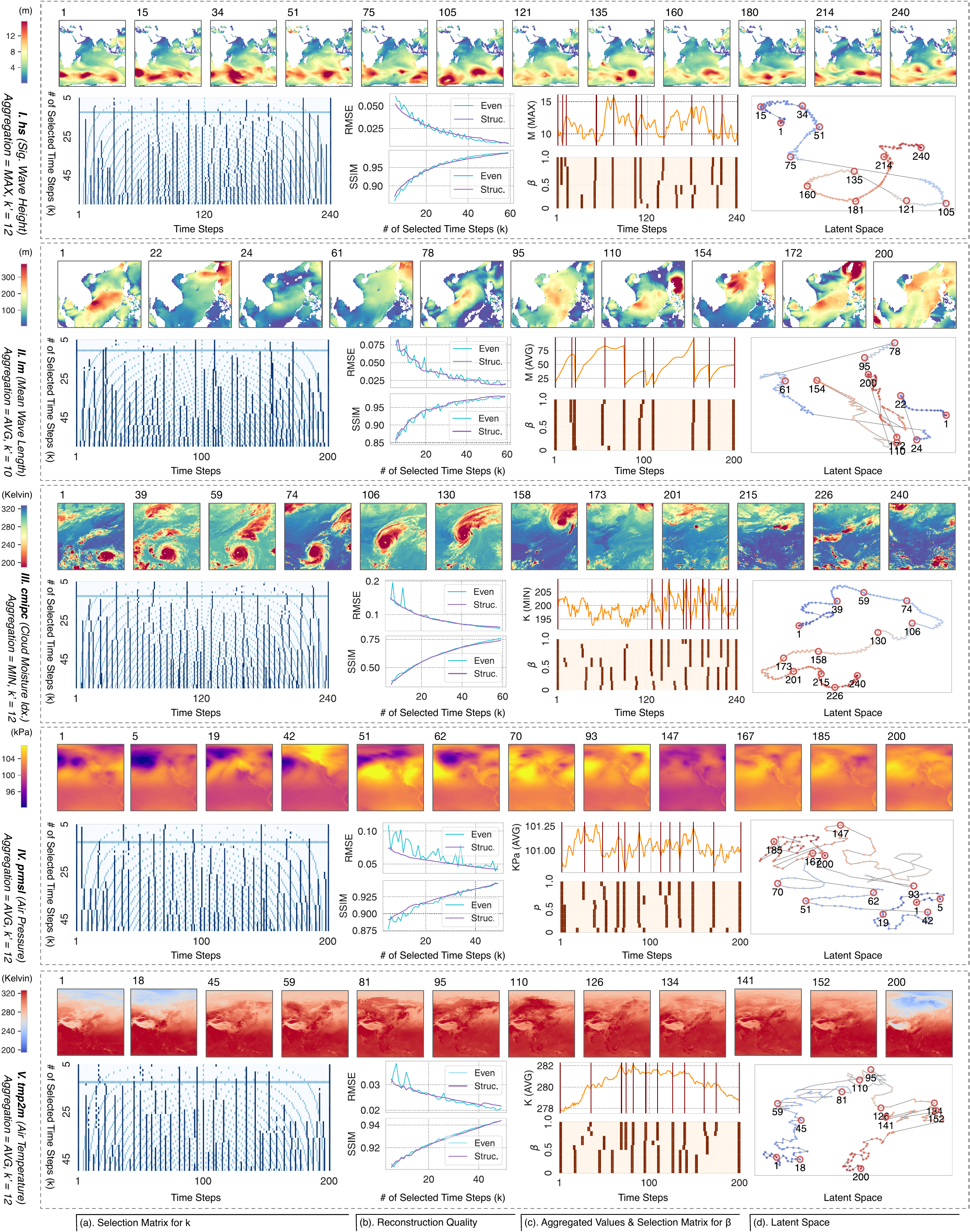}
  }
  \caption{
      Experiment Results. The meaning of the plots is detailed in Section \ref{sec:plot_meaning}.
  }
  \label{fig:exp}
  \Description{
A page-size figure with 5 rows, each row corresponding to the experiment results of one dataset. Each row contains colorized heat maps of salient frames, the selection matrix for k and \beta, PSNR and RMSE for our method and even selection, and the dimension-reduced latent space. Notably, the PSNR and RMSE for even selection tend to fluctuate as k increases, whereas our method consistently achieves better results. The heat maps displayed at the top demonstrate that our approach effectively captures significant data trends. The Latent Space trajectory on the right suggests that our method tends to select turning points as salient frames, maximizing the structural variations in the data.
  }
\end{figure*}

\autoref{fig:exp} shows the experiment results. Each row corresponds to a dataset, containing the following subfigures:

\begin{itemize}
  \item Colorized data image of the selected $k'$ salient times, juxtaposed on the top.
  \item Selection matrix for $k$. The horizontal axis represents the data's time steps, and the vertical axis represents different $k$ values. If a time step is selected under that value, the corresponding cell is colored. As a reference, results for even selection are colored in light blue.
  \item Quality of reconstructed data measured in RMSE and SSIM, where \textit{Struc.} stands for the metrics using our method, considering only the structural cost and \textit{Even} stands for the metrics using even selection.
  \item Selection matrix for $\beta$ and the aggregated data values. In this experiment, $k = k'$. The vertical lines in the aggregated data value denote the selected time steps using only the statistical cost ($\alpha = 0, \beta = 1$).
  \item Dimension-reduced Latent Space, as explained in the caption of \autoref{fig:cdis}.
\end{itemize}

% selection matrix for $k$ (a), the reconstruction quality measured in RMSE and PSNR (b), aggregated data trend and selection matrix for $\beta$ (c), and the dimension-reduced latent space (d). The colorized data for salient time steps are juxtaposed on the top of these plots. Within the selection matrices, the horizontal axis represents the time steps, and the vertical axis represents different $k$ values or $\beta$ values. If a time step is selected under that value, the corresponding cell is colored. As a reference, the evenly selection results are colored in light blue in the selection matrix for $k$. An additional row is colored with light blue in the selection matrix for $k$, denoting the selection results of the same $k$ as we highlighted in the latent space, used in the selection matrix for $\beta$, and juxtaposed on the top.

\subsection{Summarizability}

As shown in the data reconstruction quality metrics, our method's summarizability not only surpasses even selection but also exhibits greater stability. As $k$ increases, even selection often leads to erratic fluctuations in RMSE and SSIM, as observed in all 5 datasets, especially \texttt{tmp2m} and \texttt{prmsl}. In contrast, our method remains stable. This is because even selection fails to select frames that have a significant contribution to the variation of data. Similar frames may be selected, where important variations in between are omitted. In contrast, our structural cost ensures that the selected frames are structurally different from each other, thus preserving data variation.

\begin{table}[htbp]
  \sidecomment{R3.2\\MR1}
  \centering
  \small
  \caption{\revisiontext{
    Summarizability of salient time steps selected with different methods. The RMSE and SSIM are calculated between the original data and the piece-wise linear interpolated data.
  }}
  \renewcommand{\arraystretch}{0.6}
  \setlength{\tabcolsep}{3pt}
  \revision{
  \begin{tabular}{cccccccc}
  \toprule
       &      & \multicolumn{2}{c}{Ours} & \multicolumn{2}{c}{Arc-based \cite{porterDeepLearningApproach2019}} & \multicolumn{2}{c}{Even} \\
  \cmidrule(lr){3-4}\cmidrule(lr){5-6}\cmidrule(lr){7-8}Dataset & k    & RMSE $\downarrow$ & SSIM $\uparrow$ & RMSE $\downarrow$ & SSIM $\uparrow$ & RMSE $\downarrow$ & SSIM $\uparrow$ \\
  \midrule
  \multicolumn{1}{c}{\multirow{3}{*}{\texttt{tmp2m}}} & 10   & \textbf{0.0286} & \textbf{0.9127} & 0.0326 & 0.9055 & 0.0299 & 0.9111 \\
       & 20   & \textbf{0.0258} & 0.9220 & 0.0260 & \textbf{0.9234} & 0.0261 & 0.9241 \\
       & 40   & 0.0214 & 0.9403 & \textbf{0.0207} & \textbf{0.9439} & 0.0221 & 0.9430 \\
  \midrule
  \multicolumn{1}{c}{\multirow{3}{*}{\texttt{hs}}} & 10   & \textbf{0.0404} & \textbf{0.9064} & 0.0448 & 0.8998 & 0.0476 & 0.8946 \\
       & 20   & \textbf{0.0282} & \textbf{0.9430} & 0.0298 & 0.9388 & 0.0301 & 0.9413 \\
       & 40   & 0.0186 & \textbf{0.9727} & \textbf{0.0171} & 0.9720 & 0.0229 & 0.9577 \\
  \midrule
  \multicolumn{1}{c}{\multirow{3}{*}{\texttt{prmsl}}} & 10   & \textbf{0.0654} & \textbf{0.9014} & 0.0716 & 0.9006 & 0.1023 & 0.8892 \\
       & 20   & \textbf{0.0571} & 0.9143 & 0.0578 & \textbf{0.9153} & 0.0620 & 0.9135 \\
       & 40   & \textbf{0.0454} & \textbf{0.9389} & 0.0459 & 0.9365 & 0.0466 & 0.9373 \\
  \midrule
  \multicolumn{1}{c}{\multirow{3}{*}{\texttt{cmipc}}} & 10   & \textbf{0.1205} & 0.4473 & 0.1726 & 0.4459 & 0.1809 & \textbf{0.4525} \\
       & 20   & \textbf{0.0916} & \textbf{0.5745} & 0.0946 & 0.5696 & 0.0967 & 0.5737 \\
       & 40   & 0.0800 & 0.6436 & \textbf{0.0799} & \textbf{0.6546} & 0.0815 & 0.6308 \\
  \midrule
  \multicolumn{1}{c}{\multirow{3}{*}{\texttt{lm}}} & 10   & \textbf{0.0638} & \textbf{0.9004} & 0.0699 & 0.8842 & 0.0744 & 0.8799 \\
       & 20   & \textbf{0.0416} & \textbf{0.9384} & 0.0517 & 0.9175 & 0.0577 & 0.9189 \\
       & 40   & \textbf{0.0310} & 0.9574 & 0.0390 & \textbf{0.9588} & 0.0329 & 0.9624 \\
  \bottomrule
  \end{tabular}%
  }
  \label{tab:eval-arc}%
\end{table}%

\revision{
Compared with the arc-based method, our method has better reconstruction quality, particularly at small $k$ values, as listed in \autoref{tab:eval-arc}. When $k = 10$, our method has the lowest RMSE in all datasets. This indicates that our method is very effective when selecting a small number of frames. At higher $k$ values, the arc-based method may achieve slightly better quality in some cases, as in datasets \texttt{prmsl} and \texttt{lm}. This could be due to its trajectory-based selection in the dimension-reduced latent space. While losing some semantic information, it is less prone to selecting trivial frames. In contrast, our selection employs cosine similarity in the original latent space, which captures more detailed information but may suffer from data noises when selecting more frames.
}

The trajectory in the latent space further demonstrates the efficacy of our approach. Frames located at turning corners of the trajectory are likely to be selected, as these time steps typically mark the beginning or end of a particular variation. For instance, frame 74 in the \texttt{cmipc} captures the onset of the northward movement of the expanding hurricane. Frame 135 in the \texttt{hs} indicates the beginning of a decline in wave height in the southern hemisphere. As $k$ increases, these pivotal frames continue to be selected, while transitions of finer granularity are included, as illustrated in the selection matrix for $k$.

For certain $k$ values, the reconstruction quality of even selection may match or exceed our method, as seen in the \texttt{prsml} and \texttt{lm}. However, this is often followed by a sharp decline in reconstruction quality at the subsequent $k$ value. These instances occur when the even selection results overlap with the results of our method. Given that users can not preemptively determine the optimal $k$, our method offers consistent and reliable selection results.

\subsection{Anomaly and Extremum}
\label{sec:exp_anamoly}

% First: examples.

The selection matrix for $\beta$ shows the effect of the statistical cost $\mathcal{C}_{\mathrm{stat}}$. Compared with the structural cost, $\mathcal{C}_{\mathrm{stat}}$ produces more predictable results by emphasizing the changes in maximum, minimum or average values, capturing the anomalies and extremums in the datasets. This can be observed from the line chart of aggregated values when $\beta = 1.0$. For instance, when employing the \texttt{AVG} aggregation on the \texttt{tmp2m} data, several inflection points in the average values are selected. The overall temperature exhibits a trend of initial rise followed by a decline, and our method effectively selected the time steps that mark these points, together with the local fluctuations. In the \texttt{hs} data, users are concerned about the peak wave heights and by using the \texttt{MAX} aggregation, the selected time steps adeptly showcase the evolution of height extremities.

This selection method can also assist users in identifying anomalies in the data. For instance, in the \texttt{cmipc} dataset, frame 39 is selected, where the brightness temperature of this frame is abnormally low compared to surrounding frames. In the \texttt{lm} dataset, a sharp decline follows frame 154. Upon inspection, we discovered that 2 days of data were missing following this frame due to data quality issues.

Adjustable weights provide users with higher flexibility in selecting salient time steps. As shown in the selection matrix for $\beta$, when $\beta$ transitions from 0 to 1, the selection results gradually shift from prioritizing structural variation to emphasizing statistical variation. This is particularly helpful when the datasets have frames with similar spatial structures but distinct values. For example, frames 167 and 200 in the \texttt{prmsl} dataset are proximate in the latent space, indicating their similar spatial distributions. However, their average values differ by 2.4 hPa. In practice, users can specify their priorities to obtain the selections that align with their expectations.

\subsection{Runtime Performance}
\label{sec:perf}

Computing salient time steps can be divided into three stages: 1). Pre-processing: Reading the pre-stored data memory array, computing the aggregated values and cost function values in advance; 2). Latent Code Generation: Feeding the data within spatial boundaries and focus ranges to the autoencoder to obtain latent codes; 3). Frame Selection: Employing DP to select the optimal salient time steps. The time used for extracting data from NetCDF and Grib data is not considered as it needs to be executed only once for any dataset and is not perceived by users. The experiment is tested on the same server for network training as described in Section~\ref{sec:training}.

\begin{figure}[htbp]
  \centering
  \includegraphics[width=\singlecolumnwidth]{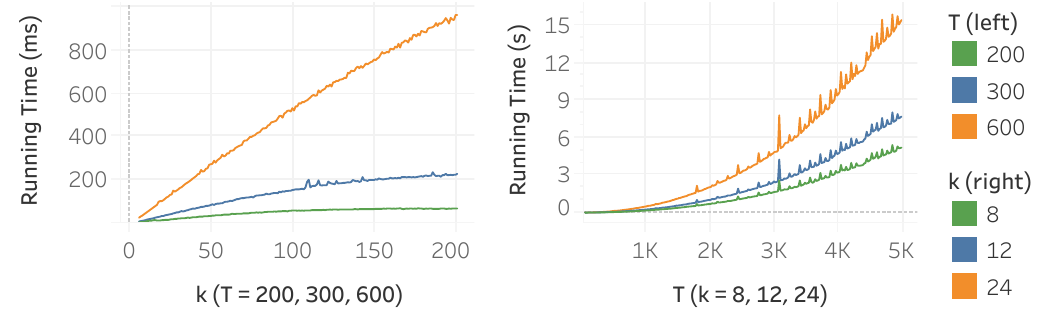}
  \caption{
      Running time of the DP process as $k$ or length of focus range $T$ increases.
  }
  \label{fig:perf}
  \Description{
    Two line charts. The left chart contains three lines, showing how running time increases as k increases, where T = 200, 300, or 600. These lines grow linearly. When k = 200, The running time for different Ts are 50ms, 200ms, and 850ms, respectively. The right chart contains three lines, showing how running time increases as T increases, where k = 8, 12, 24. These lines grow quadratically. When T  < 2k, the running time for all k values is below 2.5 seconds. When T = 5K, the running time for k = 8, 12, 24 are 5s, 7.5ss, and 15.5s, respectively.
  }
\end{figure}

\autoref{tab:perf} shows the first two stages' running time for each dataset. As shown in the table, even for datasets with thousands of frames, pre-processing and latent code generation can be completed within seconds. The running time for the third stage as $k$ or the length of focus range $T$ increases is plotted in \autoref{fig:perf}. The DP algorithm has a time complexity of $O(T^2k)$, being the dominant contributor to the total running time. We can see that the running time increases linearly with $k$ and quadratically with $T$. For cases where $T$ is under 2k frames, the selection of 24 salient time steps is completed under a second. Typically, the focus range is limited to several hundred frames, and $k$ rarely exceeds 20 frames. Under these conditions, the computational duration is below 300 ms, making it virtually instantaneous and supporting interactive selection.

\begin{table}[htbp]
  \small
  \renewcommand{\arraystretch}{0.6}
  \caption{Running time for the first two stages. Pre. stands for pre-processing. Latent. stands for latent code generation. / F stands for the average time used for processing a single frame.}
  \small
  \label{tab:perf}
  \begin{tabular}{lllll}
    \toprule
    Dataset & Pre. (s) & Pre. / F (ms) & Latent. (s) & Latent. / F (ms) \\
    \midrule
    \texttt{tmp2m} & 1.134 & 0.714 & 4.660 & 2.935 \\
    \texttt{hs} & 7.040 & 1.520 & 7.029 & 1.517 \\
    \texttt{prmsl} & 1.838 & 1.256 & 2.250 & 1.538 \\
    \texttt{cmipc} & 2.675 & 1.651 & 4.851 & 2.994 \\
    \texttt{lm} & 0.197 & 0.842 & 0.440 & 1.880 \\
    % \texttt{rdi} & Red Tide Detection Index ~\cite{shenSimpleMethodsSatellite2019} & Sentinel-3A/B & Yangtze River Estuary & 38 $\times$ 4152 $\times$ 2642 & 1 - 10 d / 1 Year \\
    \bottomrule
  \end{tabular}
\end{table}

The results of the first two stages can be cached between time selection requests, as long as the requests' spatial boundary remains unchanged. This is because the aggregation calculation depends on it. When users select a new boundary or switch datasets, they only face a few seconds of loading time before being able to perform frequent time selection interactions.

\revision{
\subsection{Expert Interview}
\sidecomment{R2.2\\R2.3\\R3.2\\R3.4\\MR3\\MR9}

To assess the usability of our system compared with existing tools mentioned in Section \ref{sec:needfinding}, we conducted an expert interview with experts in our need-finding study (E1 - E5). We employed an adapted System Usability Scale (SUS) ~\cite{Brooke1996SUSA}, a widely recognized tool that offers a comprehensive evaluation of a system's usability through 10 questions, each rated on a 5-point Likert scale. We slightly adjusted the wording to make the scale adapt to our need to focus on the data exploration process in web-based geospatial systems.

Each interview was conducted offline and lasted approximately 60 minutes. It began with a 5-minute introduction, instructing the experts to emphasize temporal navigation and selection over other system features. As there was a three-month gap since the prior study, we first let experts revisit their prior interview notes and selection tasks, allotting 10 minutes for them to familiarize themselves with these systems. Then, we let the experts engage with three systems sequentially, each for 5 to 10 minutes, followed by a SUS questionnaire for that system. After that, we introduced the features and UI components of Salientime and let the experts freely explore the system, performing identical temporal selection and download tasks as they did in the previous systems, as listed in the Outcome section of \autoref{fig:findings}. After 15 minutes, we asked them to fill out the SUS questionnaire for our system and gathered their feedback and overall opinions of our system in terms of usability and functionality.

\begin{figure*}[tbhp]
  \sidecomment{R2.3\\MR9}
  \centering
  \revisionbox{
    \includegraphics[width=\linewidth]{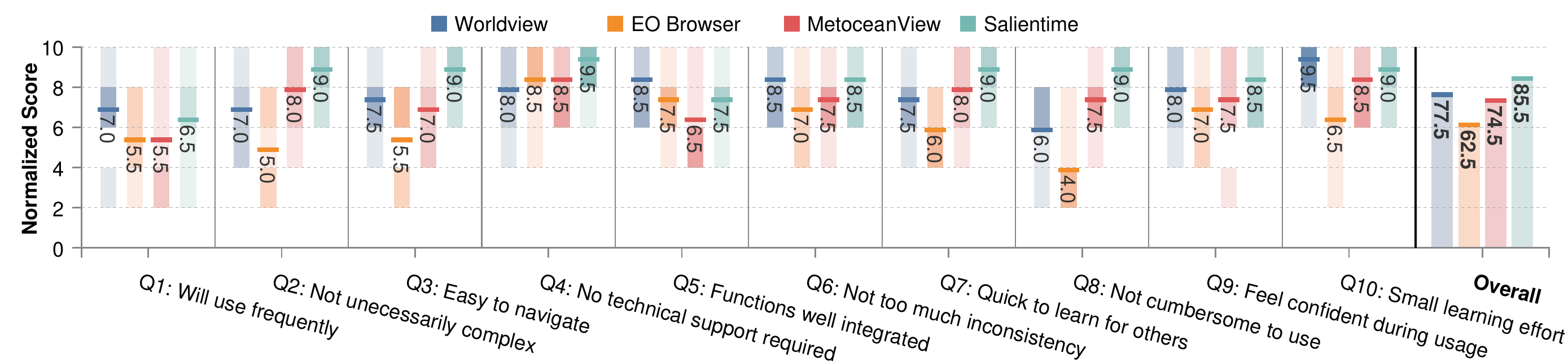}
  }
  \caption{\revisiontext{
    Usability questionnaire results. For consistency, we reverse the points for negative questions in the SUS scale and display the normalized score. The shades of the background color indicate the number of experts who selected that option.
  }}
  \label{fig:interview}
  \Description{
    A bar chart containing 11 graphs representing the scores for each question, for each system in the SUS (System Usability Scale) used in our expert interviews. The first ten sub-charts are the detailed scores for each question, and the last sub-chart is the average score for all questions. The average usability score for Worldview is 77.5, EO Browser is 62.5, MetoceanView is 74.5, and Salientime is 85.5.
  }
\end{figure*}

\autoref{fig:interview} shows the results of the questionnaire, with the usability score for each question normalized to a 0 - 10 scale. Overall, Salientime achieved a good usability score of 85.5, outperforming the other three systems (77.5, 62.5, and 74.5).

% 1). Can they effectively use the temporal context to make more informed decisions? 2). Are they able to explore both the global and regional features of the data? 3). How usable is the time-lapse video creation process? 4). Is the performance of these systems satisfactory for performing frequent temporal navigations?

% We first spent roughly 5 minutes explaining the background and that the goal is to measure the usability of different tools in terms of temporal navigation and selection, and therefore wanted the experts not to focus too much on the completeness of other features of the systems. Then, since there was a three-month gap between this interview and the previous need-finding study, we shared with experts their previous remarks and selection tasks and let them spend 10 min familiarizing themselves with the three systems. Then, we let the experts use the three systems one by one, 5 minutes each. After each use, we asked them to fill out the SUS scale for that system. After the three systems, we let the experts freely explore our system perform the same temporal selections and download tasks, and fill out the SUS scale. After that, we collected feedback about their remarks and asked questions about their overall opinions about our system.

% \autoref{fig:interview} shows results of the questionnaire. Overall, Salientime achieved a good usability score of 85.5, significantly outperforming the other three systems (77.5, 62.5, and 74.5).

% Timesteps, contextual visualization
\paragraph{Salient Frames and Temporal Context}
Experts remarked on the significance of selected salient frames in identifying crucial data variations, thus aiding decision-making about subsequent selections. Compared to existing systems that solely provide a timeline (Worldview, MetoceanView) or a date picker for inputting data (EO Browser), our timeline design that highlights salient time steps and incorporates contextual visualization offers enhanced usability in their workflows. Most experts (2 strongly agree, 3 agree, Q9) agreed that contextual visualizations on the timeline make them more confident when interacting with the system, and easier to perform temporal navigations (3 strongly agree, 2 agree, Q3).

\paragraph{Spatial Features}
The dimension-reduced latent space provides additional insight into data variation trends. As a novel approach for providing data overviews compared with existing systems, experts were particularly interested in this view and frequently inspected the frames at the turning points of the latent space trajectory. The combination of the contextual visualization on the timeline and the latent space visualization offers a comprehensive understanding of data variation and cycle information. The integration of this view with other system functions is well-received (1 strongly agree, 3 agree, Q5). Additionally, E1 and E2 predominantly utilized the regional mode to examine specific regions of the data, emphasizing the importance and usefulness of the regional selection features in geospatial systems.

% Time lapse video, creation
\paragraph{Progressive Loading and Performance}
Prioritizing the loading of salient frames and their surrounding frames improved the overall performance of our system (3 strongly agree, 2 agree, Q8). Unlike traditional methods that start frame data download only upon user navigation to a new frame, our approach significantly reduces the user-perceived latency. These salient frames also serve as the initial state for progressive time-lapse video generation. With more frames being loaded, the dynamic information becomes more accurate. This process enables users to rapidly access data dynamics, avoiding the extensive buffer loading time required by Worldview. Experts E3 and E4 recognize this as a useful engineering technique to enhance user experience.

% Learning effort, flexibility and practical usage
\paragraph{Learning Effort}
We also consulted the learning effort of our system, as we introduced extra adjustable parameters $\alpha$, $\beta$, and aggregation during selection. All experts agreed that this selection process was straightforward (3 strongly agree, 2 agree, Q7) and did not require external technical support (4 strongly agree, 1 agree). After a brief introduction to these parameters during the interview, experts can swiftly comprehend their meaning and adjust parameters to fit their needs. Moreover, E4 appreciated the flexibility provided by these parameters, remarking that the impact of different parameters is reasonable, making our system suitable for various tasks. However, E3 highlighted the need for a more detailed explanation of the temporal contexts on the timeline to ensure comprehensive understanding.

Overall, the interview result indicates a high usability of our system. With automatically selected salient frames with user-specified priorities and multiple contextual visualizations indicating data variations, our system provides easier temporal navigation, faster response time and acceptable learning effort, greatly enhancing the user experience for temporal navigation and data exploration of geospatial data.

% \paragraph{Temporal Context}~The temporal contexts provided significantly help experts in making temporal navigations. Most experts (2 strongly agree, 3 agree) think that contextual visualizations make them more confident when interacting with the system (Question 9), and easier to perform navigation (3 strongly agree, 2 agree, Question 3). E2 remarks that ``He can use the ''

% \paragraph{Global and Regional Features}~1

% \paragraph{Dynamic vs Static}~1

% \paragraph{Performace}~1
}

% Preprocessing takes time

% Running time of the DP

% \section{Expert Interview}
\section{Discussion}

\subsection{Parameter Selection}

We allow users to specify the weight for $C_{\text{struc}}$ and $C_{\text{stat}}$ in the cost function. However, the weight for distance cost is fixed. We believe the weight for the distance cost $\gamma$ should not be a user concern, as it tends to guide the selection towards even selection and should be set based on data characteristics. From our observations, $\gamma = 0.3$ serves as a general value for the distance cost. Under such value, any attempt to select two consecutive frames with a structural similarity higher than 0.7 is highly likely to be rejected. Therefore, in datasets with more drastic changes, a larger $\gamma$ value might be needed to prevent clustering of selection results, while in milder cases, $\gamma=0.1$ may suffice.

As for the dimension of the latent space $\phi$, it determines the feature extraction capability of our network. While a larger $\phi$ may offer more expressive latent codes, it also increases the risk of overfitting by learning the noise in the data. In contrast, a smaller $\phi$ may not suffice to encode the data, compromising its generalizability across geospatial datasets. Through multiple experiments, we choose $\phi = 512$ in this trade-off. Nonetheless, we believe it's worth exploring other sizes of $\phi$, depending on the volume and complexity of the dataset.

% \subsection{Alternative Design}
\subsection{Complexity and Scalability}

As discussed in Section \ref{sec:perf}, the computation is primarily divided into three stages. Both Pre-processing and Latent Code Generation have a time complexity of $O(n)$, where $n$ is the total number of frames. The DP process with a time complexity of $O(T^2k)$ is the bottleneck of our method. Although our experiments indicate fast running time in selections within 2k frames, when $T$ exceeds 5k and $k$ exceeds 20, the waiting time becomes quite lengthy. This can be seen in datasets spanning several decades or those with high temporal resolution. To scale to these situations, several strategies can be considered. One approach is to pre-compute frames with coarser temporal resolution, reducing the size of $T$ in a stepped manner. Alternatively, the data can be divided into $p$ smaller segments, and the DP process can be run in parallel on each segment, selecting $\lfloor p / k \rfloor$ frames per segment. Additionally, approximate methods of DP that sacrifice global optimality, such as the multi-pass algorithm proposed by Zhou et al. ~\cite{zhouKeyTimeSteps2018}, can be considered.

% \subsubsection{Necessarity of Distance Cost}

% \subsubsection{Max}

\subsection{Limitations and Future Work}

First, we currently focus on the raster model of geospatial data, while vector data such as wind direction or ocean currents are equally important. Meanwhile, the interrelationships between variables in multivariable geospatial data remain unexplored. Extending our approach to vector data and multivariable datasets is a potential direction for future research. Second, the DP process with a time complexity of $O(T^2k)$ poses challenges when applied to datasets with more than 5k time steps. Investigating more efficient or approximate methods could enhance the scalability of our approach. \revision{\sidecomment{R3.3\\MR2}Third, there is an inherent loss of spatial continuity when we select a subset of data frames from the full data. To further mitigate this loss, we could incorporate compression-based data transfer that enables the simultaneous transfer of surrounding frames alongside the salient frames. Visual encodings that indicate the extent of information loss can be beneficial as well.} Lastly, there's potential to incorporate more comprehensive contextual visualizations. For instance, visualizations that assist users in identifying cyclical patterns in the data, or those that combine temporal and spatial trajectories. Such tools can enable users to make more informed selections.

% 1. Multivariable & Vector Data. (Functionality)
% 2. DP Complexity.  (Efficiency)
% 3. More context visualizations. Potential cycle information. (Design)

\section{Conclusion}

In this paper, we offer a new approach and interaction workflow for time step selection in geospatial systems, aiming to enhance user experience and maximize the research value of large-scale geospatial data. Our method is built upon an extensive need-finding study with domain experts, through which we observe their time selection workflows, identify their needs, and establish a multifaceted definition for salient time steps. Leveraging autoencoders and dynamic programming, we select time steps with user-specified priorities in terms of structural and statistical variation. Moreover, we design and implement a web-based interactive system to facilitate context-aware exploration and selection of time steps. We present two case studies to showcase the efficacy of our system and evaluate the effectiveness of our methods on diverse geospatial datasets. The source code is available at \url{https://github.com/billchen2k/salientime}.

\begin{acks}
This work was supported by the NSFC under Grants (No. 61802128 and 62072183), the Shanghai Committee of Science and Technology, China (Grant No. 22511104600), and the Yangtze River Delta Science and Technology Innovation Community Project, China (Grant No. 23002400400). Chenhui Li is the corresponding author.
\end{acks}

\bibliographystyle{ACM-Reference-Format}
% \bibliography{refs, refs_overleaf}
\bibliography{main}
% \appendix

\end{document}